\pdfoutput=1
\documentclass[prd,showpacs,letterpaper,10pt,aps,notitlepage,nofootinbib,superscriptaddress]{revtex4-2}

\usepackage{amsfonts,amsmath,graphicx}
\usepackage{hyperref}
\usepackage{color}
\usepackage{placeins}
\usepackage{diagbox}
\usepackage[capitalize]{cleveref}
\usepackage{bm}

\usepackage{xargs}                    % Use more than one optional parameter in a new commands
\usepackage[pdftex,dvipsnames]{xcolor} 
\usepackage[colorinlistoftodos,prependcaption,textsize=small]{todonotes}

\hypersetup{colorlinks, linkcolor = [rgb]{0,0.0,0.75}, citecolor = [rgb]{0,0.0,0.75}, urlcolor = [rgb]{0,0.0,0.75}}

\newcommand{\C}{\mathcal{C}}

\newcommand{\xb}{\textbf{x}}
\newcommand{\yb}{\textbf{y}}
\newcommand{\pb}{\textbf{p}}
\newcommand{\dagg}{^\dagger}
\newcommand{\op}{\mathcal{O}}
\newcommand{\vp}{\phantom{+}}

\newcommand{\vz}{\phantom{0}}

\newcommand{\bbud}{\bar{b}\bar{b}ud}
\newcommand{\bbus}{\bar{b}\bar{b}us}

\newcommandx{\unsure}[2][1=]{\todo[linecolor=red,backgroundcolor=red!25,bordercolor=red,#1]{#2}}
\newcommandx{\change}[2][1=]{\todo[linecolor=blue,backgroundcolor=blue!25,bordercolor=blue,#1]{#2}}
\newcommandx{\missing}[2][1=]{\todo[linecolor=Orange,backgroundcolor=Orange!25,bordercolor=Orange,#1]{#2}}
\newcommandx{\info}[2][1=]{\todo[linecolor=OliveGreen,backgroundcolor=OliveGreen!25,bordercolor=OliveGreen,#1]{#2}}
\newcommandx{\improve}[2][1=]{\todo[linecolor=Plum,backgroundcolor=Plum!25,bordercolor=Plum,#1]{#2}}

\newcolumntype{C}[1]{>{\centering\arraybackslash}m{#1}}

\definecolor{darkgreen}{rgb}{0,0.5,0}

\begin{document}

\title{ \texorpdfstring{$\bm{\bar b \bar b u d}$}{bbud} and \texorpdfstring{$\bm{\bar b \bar b u s}$}{bbus} tetraquarks from lattice QCD using symmetric correlation matrices with both local and scattering interpolating operators}

\author{Constantia Alexandrou}
\affiliation{Department of Physics, University of Cyprus, 20537 Nicosia, Cyprus}
\affiliation{Computation-based Science and Technology Research Center, The Cyprus Institute, 20 Konstantinou Kavafi Street, 2121 Nicosia, Cyprus}

\author{Jacob Finkenrath}
\affiliation{Bergische Universit\"at Wuppertal, Gau{\ss}stra{\ss}e 20, D-42119 Wuppertal, Germany}

\author{Theodoros Leontiou}
\affiliation{Department of Mechanical Engineering, Frederick University, 1036 Nicosia, Cyprus}

\author{Stefan Meinel}
\affiliation{Department of Physics, University of Arizona, Tucson, AZ 85721, USA}

\author{Martin Pflaumer}
\affiliation{Goethe-Universit\"at Frankfurt am Main, Institut f\"ur Theoretische Physik, Max-von-Laue-Stra{\ss}e 1, D-60438 Frankfurt am Main, Germany}

\author{Marc Wagner}
\affiliation{Goethe-Universit\"at Frankfurt am Main, Institut f\"ur Theoretische Physik, Max-von-Laue-Stra{\ss}e 1, D-60438 Frankfurt am Main, Germany}
\affiliation{Helmholtz Research Academy Hesse for FAIR, Campus Riedberg, Max-von-Laue-Stra{\ss}e 12, D-60438 Frankfurt am Main, Germany}

\date{April 4, 2024}

\begin{abstract}
We study the $\bar b \bar b u d$ tetraquark with quantum numbers $I(J^P) = 0(1^+)$ as well as the $\bar b \bar b u s$ tetraquark with quantum numbers $J^P = 1^+$ using lattice QCD. We improve on existing work by including both local and scattering interpolating operators on both sides of the correlation functions and use symmetric correlation matrices. This allows not only a reliable determination of the energies of QCD-stable tetraquark ground states, but also of low-lying excited states, which are meson-meson scattering states. The latter is particularly important for future finite-volume scattering analyses. Here, we perform chiral and continuum extrapolations of just the ground-state energies, for which finite-volume effects are expected to be small. Our resulting tetraquark binding energies, $-100 \pm 10\:^{+36}_{-51}\:\:{\rm MeV}$ for $\bar b \bar b u d$ and $-30 \pm 3\:^{+11}_{-31}\:\:{\rm MeV}$ for $\bar b \bar b u s$, are consistent with other recent lattice-QCD predictions.
\end{abstract}

\maketitle

% ********************
% ********************
% ********************
% ********************
% ********************

\section{\label{sec:intro}Introduction}

Mesons, which are hadrons with integer spin, are typically composed of one valence quark and one valence antiquark. However, they can as well contain two valence quarks and two valence antiquarks. Such exotic states are called tetraquarks\footnote{In the literature, the term ``tetraquark'' is somewhat ambiguous. In certain papers it exclusively refers to a diquark-antidiquark structure, while in other papers it is used more generally for arbitrary bound states and resonances with a strong four-quark component, including, e.g., mesonic molecules. Throughout this paper we follow the latter convention.}. In this work we use lattice QCD to study two particular antiheavy-antiheavy-light-light four-quark systems, the $\bar b \bar b u d$ tetraquark with quantum numbers $I(J^P) = 0(1^+)$ and the $\bar b \bar b u s$ tetraquark with quantum numbers $J^P = 1^+$. These tetraquarks are expected to be QCD-stable (i.e., have a mass lower than the sum of the masses of the lightest possible strong-decay products, a pseudoscalar and a vector heavy-light meson in this case) for sufficiently large heavy-quark mass \cite{Carlson:1987hh,Manohar:1992nd,Eichten:2017ffp}. Accurately predicting the binding energies at the physical $b$-quark mass is quite challenging. Approaches using potential models, effective field theories, and QCD sum rules \cite{Carlson:1987hh,Manohar:1992nd,SilvestreBrac:1993zem,Brink:1998as,Vijande:2003ki,Janc:2004qn,Vijande:2006jf,Navarra:2007yw,Ebert:2007rn,Zhang:2007mu,Lee:2009rt,Karliner:2017qjm,Eichten:2017ffp,Wang:2017uld,Park:2018wjk,Deng:2018kly,Wang:2018atz,Liu:2019stu,Tan:2020ldi,Lu:2020rog,Braaten:2020nwp,Faustov:2021hjs,Guo:2021yws,Dai:2022ulk,Kim:2022mpa,Chen:2022ros,Praszalowicz:2022sqx,Richard:2022fdc,Wu:2022gie,Maiani:2022qze,Song:2023izj,Liu:2023vrk} gave a wide spread of results for the binding energy of the $\bar b \bar b u d$ system, and did not all predict the $\bar b \bar b u s$ mass to be below threshold (see, e.g., the summary plots in Fig.~\ref{FIG:compare} in the conclusions section of this paper). On the other hand, studies using lattice QCD consistently find both the $\bar b \bar b u d$ and $\bar b \bar b u s$ tetraquarks to be QCD-stable. While early investigations based on static potentials from lattice QCD and the Born-Oppenheimer approximation slightly underestimated the binding energy of the $\bar b \bar b u d$ tetraquark \cite{Bicudo:2012qt,Brown:2012tm,Bicudo:2015kna,Bicudo:2015vta,Bicudo:2016ooe}, more recent rigorous full lattice-QCD simulations \cite{Francis:2016hui,Junnarkar:2018twb,Leskovec:2019ioa,Mohanta:2020eed,Meinel:2022lzo,Hudspith:2023loy,Aoki:2023nzp} obtained binding energies of $\mathcal{O}(100 \, \text{MeV})$ for the $\bar b \bar b u d$ tetraquark with $I(J^P) = 0(1^+)$ and $\mathcal{O}(50 \, \text{MeV})$ for the $\bar b \bar b u s$ tetraquark with $J^P = 1^+$, with good agreement within statistical and systematic uncertainties among the more advanced calculations.

There are further antiheavy-antiheavy-light-light systems that are promising with respect to the existence of tetraquark bound states or resonances, but for whom the situation is less clear. Two such candidates are the $\bar b \bar c u d$ systems with quantum numbers $I(J^P) = 0(0^+)$ and $I(J^P) = 0(1^+)$. Independent groups carrying out full lattice-QCD simulations have arrived at different conclusions. The existence of a strong-interaction-stable $\bar b \bar c u d$ tetraquark with $I(J^P) = 0(1^+)$ was initially reported in Ref.\ \cite{Francis:2018jyb} but later revoked in Ref.~\cite{Hudspith:2020tdf}. In Ref.~\cite{Meinel:2022lzo}, we did not find evidence for QCD-stable $\bar b \bar c u d$ bound states, but could not \emph{rule out} shallow bound states. Other authors found an indication for a $\bar b \bar c u d$ bound state with $I(J^P) = 0(1^+)$ below threshold \cite{Mathur:2021gqn,Padmanath:2023rdu}. The conclusions of these studies were based only on the finite-volume ground-state energy. In a recent project carried out in parallel to this work we revisited both $\bar b \bar c u d$ systems taking into account also finite-volume energies of low-lying excitations. We found strong indication for very shallow bound states \cite{Alexandrou:2023cqg}.

Another interesting system is $\bar b \bar b u d$ with quantum numbers $I(J^P) = 0(1^-)$. This system has not yet been investigated in full lattice QCD, but has been explored in the Born-Oppenheimer approximation using lattice-QCD static potentials. While, at first, a tetraquark resonance was predicted in a crude way (completely neglecting effects of the heavy-quark spins) \cite{Bicudo:2017szl}, a refined study that is still ongoing suggests that neither a bound state nor a resonance exist in the region near the $B B$ threshold \cite{Hoffmann:2022jdx}. Given that the Born-Oppenheimer approximation with lattice-QCD static potentials seems to underestimate the binding energy of QCD-stable antiheavy-antiheavy-light-light systems, the existence of a $I(J^P) = 0(1^-)$, $\bar b \bar b u d$ resonance is not yet excluded and should be investigated in full lattice QCD.

To rigorously search for tetraquark resonances or shallow bound states in full lattice QCD, determinations of the meson-meson scattering amplitudes both below and above the relevant thresholds are mandatory. Such computations are typically based on L\"uscher's method \cite{Luscher:1990ux} and its generalizations (see the review in Ref.~\cite{Briceno:2017max}), and require a precise computation of all relevant low-lying finite-volume energy levels. To be able to resolve all of these energy levels, one needs a sufficiently large basis of interpolating operators that must be able to capture the spatial structures of all relevant states. This requires the use of both local operators, in which the four quarks are placed at or near the same spatial point and jointly projected to the desired total momentum, and scattering operators, which are constructed from products of two individually momentum-projected meson operators.

Previous full lattice-QCD studies of antiheavy-antiheavy-light-light systems containing anti-bottom quarks used either exclusively local operators \cite{Francis:2016hui,Junnarkar:2018twb,Leskovec:2019ioa,Mohanta:2020eed,Hudspith:2023loy,Aoki:2023nzp,Padmanath:2023rdu} or a combination of local and scattering operators with the latter employed only at the sink of the correlation functions \cite{Leskovec:2019ioa,Meinel:2022lzo}.
In this paper (and in our very recent work \cite{Alexandrou:2023cqg} on the $\bar b \bar c u d$ system, which we carried out in parallel) we implemented and used, for the first time in the bottom sector, local and scattering interpolating operators both at the source and at the sink of correlation functions, leading to symmetric correlation matrices containing both types of operators. Here we focus again on the $\bar b \bar b u d$ system with quantum numbers $I(J^P) = 0(1^+)$ and the $\bar b \bar b u s$ system with quantum numbers $J^P = 1^+$. We demonstrate that we are able to extract the energy levels of several meson-meson scattering states in addition to the ground-state energy below the lowest meson-meson threshold (however, in contrast to Ref.~\cite{Alexandrou:2023cqg}, here we only included scattering operators with vanishing relative momenta and were therefore only able to resolve the subset of scattering states of that type).

We use a lattice setup quite different from our previous lattice-QCD studies \cite{Leskovec:2019ioa,Meinel:2022lzo} of antiheavy-antiheavy-light-light systems. The reason is that in our previous work we could reuse existing domain-wall point-to-all propagators previously generated for other projects \cite{Meinel:2016dqj,Meinel:2020owd,Meinel:2021mdj,Meinel:2021rbm}. However, the computation of symmetric correlation matrices with both local and scattering operators, i.e.\ with the same set of local and scattering operators used at the source as well as at the sink, is not possible with only point-to-all propagators. One additionally needs stochastic timeslice-to-all propagators for the correlation matrix elements with scattering operators at the source. Technical details can be found, for example, in Ref.\ \cite{Abdel-Rehim:2017dok}, which includes a discussion of the one-end-trick that is essential for computing matrix elements with stochastic propagators. Since we had to recompute light and strange quark propagators, we decided to use a computationally cheaper mixed-action setup previously employed by the PNDME collaboration \cite{Bhattacharya:2015wna,Gupta:2018qil}: clover-improved Wilson  (with HYP link smearing) valence $u$, $d$, and $s$ quarks on the $(2+1+1)$-flavor gauge-link ensembles generated by the MILC collaboration using the highly improved staggered quark (HISQ) action \cite{MILC:2012znn}. As before, we use lattice NRQCD for the heavy $\bar b$ quarks.

While it is certainly interesting to revisit the $\bar b \bar b u d$ system with quantum numbers $I(J^P) = 0(1^+)$ and the $\bar b \bar b u s$ system with quantum numbers $J^P = 1^+$, now for the first time with a combination of local and scattering operators and symmetric correlation matrices, the main motivation of this work is rather to explore and prepare computational methods suited to study antiheavy-antiheavy-light-light tetraquark \emph{resonances}. As discussed above, this requires a precise determination of energy levels of low-lying scattering states. We demonstrate that this works well with our setup, and also present evidence that it is not feasible without scattering operators.

This article is organized in the following way. In \cref{sec:latticesetup} we briefly summarize our lattice setup. In \cref{SEC597} we discuss the interpolating operators for the two systems we investigate and the corresponding correlation functions. In \cref{SEC332} we provide our lattice results for the energies of the pseudoscalar and vector $B$ and $B_s$ mesons and determine their kinetic masses from the momentum dependence of these energies. In \cref{sec:results} we present the numerical results for the low-lying finite-volume energy levels of the $\bar{b} \bar{b} u d$ and $\bar{b} \bar{b} u s$ four-quark systems. We also explore the importance of each of our interpolating operators, which provides certain insights concerning the structure of the low-lying states. In \cref{sec:FVeffects} we comment on finite-volume effects and the left-hand cut issue.
Our extrapolations of the binding energies to the physical pion mass and continuum limit are discussed in \cref{sec:extrap}.
Finally, we summarize the main points of our work in \cref{SEC596} and give a brief outlook. Note that results obtained at an early stage of this project were presented at the Lattice 2022 conference \cite{Wagner:2022bff}.

% ********************
% ********************
% ********************
% ********************
% ********************

\FloatBarrier

\section{\label{sec:latticesetup}Lattice setup}

We use $(2+1+1)$-flavor gauge-link ensembles generated by the MILC collaboration using the highly improved staggered quark (HISQ) action and a one-loop Symanzik improved gauge action (see Ref.\ \cite{MILC:2012znn} for details). The main properties of the seven ensembles, which differ in the lattice spacing, the spatial volume, and the pion mass, are collected in Table~\ref{tab:configurations}. The relative lattice spacings were determined using $r_1/a$, which is related to the static quark-antiquark force $F(r)$ and defined via $r_1^2 F(r_1) = 1$; the physical value $r_1 = 0.3106(8)(14)(4) \, \text{fm}$ was obtained through an analysis of pseudoscalar decay constants \cite{MILC:2012znn,MILC:2010hzw}. We note that the ensembles a12m220S, a12m220 and a12m220L have the same gauge coupling and bare light quark mass and only differ in the number of lattice sites in the spatial directions. Thus, they are particularly suited to investigate the volume dependence of energy eigenvalues.

\begin{table}[h]
	\centering
	\begin{tabular}{cccccc}\hline \hline
		ensemble & $ N_s^3\times N_t $ & $ a $ [fm] & $ m_{\pi}^{(\textrm{sea})} $ [MeV] & $ m_{\pi}^{(\text{val})} $ [MeV] & $ N_{\text{conf}} $ \\
		\hline
		a15m310  & $ 16^3 \times 48 $ & 0.1510(20) & $ 306.9(5) $ & $ 320.6(4.3) $ & $ 11554 $\\
		\hline
		a12m310  & $ 24^3 \times 64 $ & 0.1207(11) & $ 305.3(4) $ & $ 310.2(2.8) $ & $ \phantom{0}1053 $\\
		a12m220S & $ 24^3 \times 64 $ & 0.1202(12) & $ 218.1(4) $ & $ 225.0(2.3) $ & $ \phantom{0}1020 $\\
		a12m220  & $ 32^3 \times 64 $ & 0.1184(10) & $ 216.9(2) $ & $ 227.9(1.9) $ & $ \phantom{0}1000 $\\
		a12m220L & $ 40^3 \times 64 $ & 0.1189(09) & $ 217.0(2) $ & $ 227.6(1.7) $ & $ \phantom{0}1030 $\\
		\hline
		a09m310  & $ 32^3 \times 96 $ & 0.0888(08) & $ 312.7(6) $ & $ 313.0(2.8) $ & $ \phantom{0}1166 $\\
		a09m220  & $ 48^3 \times 96 $ & 0.0872(07) & $ 220.3(2) $ & $ 225.9(1.8) $ & $ \phantom{00}657 $\\ \hline \hline& 
	\end{tabular}
	\caption{\label{tab:configurations}Gauge-link ensembles used in this work. Here, $N_s$, $N_t$ are the numbers of lattice sites in spatial and temporal directions, $a$ is lattice spacing, $m_{\pi}^{(\rm sea)}$ is the HISQ sea-quark pion mass (taste $\gamma_5$), and $N_{\text{conf}}$ is the number of gauge-link configurations \cite{MILC:2012znn}.
	Also shown are the results for the valence pion masses $m_{\pi}^{(\rm val)}$ computed on these ensembles using the Wilson-clover action with the parameters in Table \protect\ref{tab:Cloverparams_MILC} \cite{Bhattacharya:2015wna,Gupta:2018qil}.}
\end{table}

For the valence $u$, $d$, and $s$ quarks, we use the Wilson-clover action with HYP-smeared gauge links to suppress exceptional configurations that can appear in a mixed-action setup (see the detailed discussion in Section~II.A in Ref.\ \cite{Bhattacharya:2015wna}). This mixed-action setup was tested and successfully used by the PNDME collaboration in the context of computations of isovector and isoscalar tensor charges of the nucleon \cite{Bhattacharya:2015wna,Gupta:2018qil}. We use the values for the bare quark masses and the Sheikholeslami-Wohlert coefficient tuned and provided by the PNDME collaboration (see Table~\ref{tab:Cloverparams_MILC} as well as Table~II of Ref.\ \cite{Bhattacharya:2015wna} and Table~I and Table~II of Ref.\ \cite{Gupta:2018qil}). The corresponding valence-quark pion masses are listed in Table~\ref{tab:configurations} and quite similar to their counterparts computed within the unitary setup. A similar statement holds for the kaon masses, where the $s$-quark mass is close to the physical $s$-quark mass for all seven ensembles.

\begin{table}[h]
	\centering
	\begin{tabular}{ccccccccccccccc}
		\hline\hline
		ensemble & \hspace{2ex} &  $a m_l^{(\text{val})}$  & \hspace{2ex} & $a m_s^{(\text{val})}$ & \hspace{2ex} & $c_{\textrm{sw}}$& \hspace{2ex} &  $a m_b$  & \hspace{2ex} & $u_{0L}$ & \hspace{2ex} & $c_1$ & $ c_5 $ & $ c_6 $ \\
		\hline
		a15m310         &&  $ -0.0893\vz $  &&  $ -0.021\vz\vz\vz $ &&  $ 1.05945 $ &&  $ 3.42 $  &&  $ 0.8195  $ &&  $ 1.36 $ & $ 1.21 $ & $ 1.36 $ \\
		a12m310         &&  $ -0.0695\vz $  &&  $ -0.018718 $  &&  $ 1.05094 $ &&  $ 2.66 $  &&  $ 0.834\vz $  &&  $ 1.31 $ & $ 1.16 $ & $ 1.31 $ \\
		a12m220S, a12m220, a12m220L	&&  $ -0.075\vz\vz $  &&  $ -0.02118\vz$  &&  $ 1.05091 $ &&  $ 2.62 $  &&  $ 0.8349 $  &&  $ 1.31 $ & $ 1.16 $ & $ 1.31 $ \\
		a09m310         &&  $ -0.05138 $  &&  $-0.016075 $  &&  $ 1.04243 $ &&  $ 1.91 $  &&  $ 0.8525 $  &&  $ 1.21 $ & $ 1.12 $ & $ 1.21 $ \\
		a09m220         &&  $ -0.0554\vz $  &&  $ -0.01761\vz $  &&  $ 1.04239 $ &&  $ 1.90 $  &&  $ 0.8521 $  &&  $ 1.21 $ & $ 1.12 $ & $ 1.21 $ \\
		\hline\hline
	\end{tabular}
	\caption{\label{tab:Cloverparams_MILC}Parameters of the Wilson-clover and NRQCD valence-quark actions. Here, $am_l^{(\text{val})}$ and $am_s^{(\text{val})}$ are the bare light and strange quark masses and $c_{\textrm{sw}}$ is the Sheikholeslami-Wohlert coefficient \cite{Bhattacharya:2015wna,Gupta:2018qil}. In the NRQCD action, $am_b$ is the bare $b$-quark mass, $u_{0L}$ is the Landau-gauge mean link used for tadpole improvement, and $c_1$, $c_5$, $c_6$ are matching coefficients of kinetic terms (see Ref.\ \cite{HPQCD:2011qwj} for details).}
\end{table}

The $b$ quarks appear only as valence quarks and are implemented using lattice NRQCD \cite{Lepage:1992tx}. We use the same order-$v^4$ action as in Ref.\ \cite{HPQCD:2011qwj}. In contrast to the light and strange valence-quark action, the gauge links in the NRQCD action are not smeared. The matching coefficients of the kinetic terms, $c_1$, $c_5$, and $c_6$, include one-loop radiative corrections, while the other matching coefficients are set to their tree-level values. All parameters were taken from Tables~II and III in Ref.\ \cite{HPQCD:2011qwj} and are also reproduced here in Table~\ref{tab:Cloverparams_MILC}.

% ********************
% ********************
% ********************
% ********************
% ********************

\FloatBarrier

\section{\label{SEC597}Interpolating operators and correlation functions}

% ********************
% ********************
% ********************

\subsection{\label{sec:fourQuarkSystems}$ \bar{b} \bar{b} u d $ and $ \bar{b}\bar{b}us $ four-quark systems}

% **********

\subsubsection{\label{sec:ops_bbud}Interpolating operators for $ \bar{b} \bar{b} u d $ with $ I(J^P)=0(1^+) $}

For the $ \bar{b} \bar{b} u d $ case, we use exactly the same interpolating operators as in our previous work \cite{Leskovec:2019ioa} (but now compute all $5\times5$ correlation matrix elements): three local operators
\begin{align}
\label{eq:op_BastB_total_zero} &\op_1 = \op_{[B B^\ast](0)} = \frac{1}{\sqrt{V_s}} \sum_{\xb}	\bar{b}\gamma_j u(\xb) \, \bar{b}\gamma_5 d(\xb) - (u \leftrightarrow d), \\
\label{eq:op_BastBast_total_zero} &\op_2 = \op_{[B^\ast B^\ast](0)} = \epsilon_{j k l} \frac{1}{\sqrt{V_s}} \sum_{\xb} \bar{b}\gamma_k u(\xb) \, \bar{b}\gamma_l d(\xb) - (u \leftrightarrow d), \\
\label{eq:op_Dd_total_zero}	&\op_3 = \op_{[D d](0)} = \frac{1}{\sqrt{V_s}} \sum_{\xb}\bar{b}^a \gamma_j \C \bar{b}^{b,T}(\xb)\, u^{a,T}  \C \gamma_5  d^b(\xb) - (u \leftrightarrow d),
\end{align}
and two scattering operators
\begin{align}
	\label{eq:op_BastB_sep_zero} &\op_4 = \op_{B(0) B^\ast(0)} =\bigg(\frac{1}{\sqrt{V_s}} \sum_{\xb}	\bar{b}\gamma_j u(\xb)\bigg) \, \bigg(\frac{1}{\sqrt{V_s}} \sum_{\yb}\bar{b}\gamma_5 d(\yb)\bigg)	- (u \leftrightarrow d), \\
	 &\op_5 = \op_{B^\ast(0) B^\ast(0)} = \epsilon_{j k l} \bigg(\frac{1}{\sqrt{V_s}} \sum_{\xb}	\bar{b}\gamma_k u(\xb)\bigg) \, \bigg(\frac{1}{\sqrt{V_s}} \sum_{\yb}\bar{b}\gamma_l d(\yb)\bigg) - (u \leftrightarrow d)
	\label{eq:op_BastBast_sep_zero} 
\end{align}
($ a,b $ are color indices, $ j,k,l $ are spatial indices, $ \C= \gamma_0 \gamma_2 $ denotes the charge conjugation matrix, and $V_s$ is the spatial volume). Both the local and the scattering operators contain the meson-meson combinations $B B^\ast$ and $B^\ast B^\ast$. The former is an obvious choice, because $B B^\ast$ represents the lowest meson-meson threshold in the $ \bar{b} \bar{b} u d $ and $ I(J^P)=0(1^+) $ sector. The $B^\ast B^\ast$ structure is also expected to be important. In particular, the local operator $\op_{[B^\ast B^\ast](0)}$ might generate a sizable overlap to the $ \bar{b} \bar{b} u d $ tetraquark as indicated by the results of Ref.\ \cite{Bicudo:2016ooe}, where the same system was investigated using lattice-QCD static potentials and the Born-Oppenheimer approximation. In addition to these four meson-meson operators, we also use a local operator of diquark-antidiquark type. The latter is motivated, for example, by Ref.~\cite{Bicudo:2021qxj}, which suggests that the $ \bar{b} \bar{b} u d $ tetraquark is an approximately even mix of a meson-meson component and a diquark-antidiquark component. For a more detailed discussion of these operators we refer to Ref.\ \cite{Leskovec:2019ioa}.

% **********

\subsubsection{\label{sec:fourQuarkSystems_bbus}Interpolating operators for $ \bar{b}\bar{b}us $ with $ J^P=1^+ $}

For the $ \bar{b} \bar{b} u s $ case we also follow our previous work \cite{Meinel:2022lzo} and use seven interpolating operators (which are now included in any combination at source and sink): four local operators
\begin{align}
\label{eq:op_BBsast_total_zero} &\op_1 = \op_{[B B_s^\ast](0)} = \frac{1}{\sqrt{V_s}} \sum_{\xb}	\bar{b}\gamma_5 u(\xb) \, \bar{b}\gamma_j s(\xb),	 \\
\label{eq:op_BastBs_total_zero} &\op_2 = \op_{[B^\ast B_s](0)} = \frac{1}{\sqrt{V_s}} \sum_{\xb}	\bar{b}\gamma_j u(\xb) \, \bar{b}\gamma_5 s(\xb),	 \\
\label{eq:op_BastBsast_total_zero} &\op_3 = \op_{[B^\ast B_s^\ast](0)} =  \epsilon_{j k l} \frac{1}{\sqrt{V_s}} \sum_{\xb}	\bar{b}\gamma_k u(\xb) \, \bar{b}\gamma_l s(\xb), \\
\label{eq:op_Dds_total_zero}&\op_4 = \op_{[D d](0)} = \frac{1}{\sqrt{V_s}} \sum_{\xb}\bar{b}^a \gamma_j \C \bar{b}^{b,T}(\xb)\, u^{a,T}  \C \gamma_5  s^b(\xb),
\end{align}
and three scattering operators
\begin{align}
\label{eq:op_BBsast_sep_zero} &\op_5 = \op_{B(0) B_s^\ast(0)} = \bigg(\frac{1}{\sqrt{V_s}} \sum_{\xb}	\bar{b}\gamma_5 u(\xb)\bigg) \, \bigg(\frac{1}{\sqrt{V_s}} \sum_{\yb}\bar{b}\gamma_j s(\yb)\bigg), \\
\label{eq:op_BastBs_sep_zero} &\op_6 = \op_{B^\ast(0) B_s(0)} = \bigg(\frac{1}{\sqrt{V_s}} \sum_{\xb}	\bar{b}\gamma_j u(\xb)\bigg) \, \bigg(\frac{1}{\sqrt{V_s}} \sum_{\yb} \bar{b}\gamma_5 s(\yb)\bigg), \\
\label{eq:op_BastBsast_sep_zero} &\op_7 = \op_{B^\ast(0) B_s^\ast(0)} = \epsilon_{j k l} \bigg(\frac{1}{\sqrt{V_s}} \sum_{\xb}	\bar{b}\gamma_k u(\xb)\bigg) \, \bigg(\frac{1}{\sqrt{V_s}} \sum_{\yb} \bar{b}\gamma_l s(\yb)\bigg) .
\end{align}
The operators (\ref{eq:op_BBsast_total_zero}) to (\ref{eq:op_BastBsast_sep_zero}) are a generalization of the operators (\ref{eq:op_BastB_total_zero}) to (\ref{eq:op_BastBast_sep_zero}) from the case with mass-degenerate $u$ and $d$ quarks to the unequal-mass case of $u$ and $s$ quarks. Specifically, the operators (\ref{eq:op_BastBsast_total_zero}), (\ref{eq:op_Dds_total_zero}) and (\ref{eq:op_BastBsast_sep_zero}) are the counterparts of the operators (\ref{eq:op_BastBast_total_zero}), (\ref{eq:op_Dd_total_zero}) and (\ref{eq:op_BastBast_sep_zero}). The main difference is due to the isospin symmetry of the $u$ and the $d$ quark, where for $I = 0$ only the antisymmetric combination $ud - du$ is allowed. In contrast, for the $u$ and the $s$ quark there is no such restriction, leading to two possibilities when combining a pseudoscalar and a vector meson: the operators (\ref{eq:op_BBsast_total_zero}) and (\ref{eq:op_BastBs_total_zero}) correspond to operator (\ref{eq:op_BastB_total_zero}) and the operators (\ref{eq:op_BBsast_sep_zero}) and (\ref{eq:op_BastBs_sep_zero}) correspond to operator (\ref{eq:op_BastB_sep_zero}). In terms of meson pairs, the two $B B^\ast$ operators in the $ \bar{b} \bar{b} u d $ case correspond to the two $B B_s^\ast$ and the two $B^\ast B_s$ operators in the $ \bar{b} \bar{b} u s $ case.

\subsubsection{Quark field and gauge link smearing}

To enhance the overlaps with the low-lying states of interest, we apply standard smearing techniques in the interpolating operators defined in \cref{sec:ops_bbud} and \cref{sec:fourQuarkSystems_bbus}. The $u$, $d$, $s$, and $b$ quark fields are Gaussian smeared (see, e.g., Eq.\ (8) in Ref.\ \cite{Leskovec:2019ioa}) with parameters listed in Table~\ref{tab:MILC_smearingParams}. Here, $\sigma_\text{Gauss}$ represents the Gaussian width of the smeared quark field in units of the lattice spacing. For the light and strange quarks, these values were chosen such that the smearing width in physical units, $\sigma_\text{Gauss} a$, is essentially independent of the lattice spacing ($\sigma_\text{Gauss} a \approx 0.55 \, \text{fm}$ for the $u$ and $d$ quarks, $\sigma_\text{Gauss} a \approx 0.45 \, \text{fm}$ for the $s$ quark). The gauge links needed for Gaussian smearing are APE smeared (see, e.g., Eq.\ (23) in Ref.\ \cite{Jansen:2008si}) with parameters $N_\textrm{APE} = 50$ and $\alpha_\textrm{APE} = 0.5$.

\begin{table}[h]
	\centering

	\begin{tabular}{ccccccccc} \hline \hline
		ensemble & \multicolumn{2}{c}{$u$ and $d$ quarks} && \multicolumn{2}{c}{$s$ quarks} && \multicolumn{2}{c}{$b$ quarks} \\ 
		& $N_\textrm{Gauss}$ & $\sigma_\textrm{Gauss}$ &\hspace{2ex}	% up,down
		& $N_\textrm{Gauss}$ & $\sigma_\textrm{Gauss}$ &\hspace{2ex}			% strange
		& $N_\textrm{Gauss}$ & $\sigma_\textrm{Gauss}$   		% bottom
		\\ \hline						
		a15m310		& $22$ & $3.55$ && $ 15 $ & $2.93$ && $ 10 $ & $ 1.0 $  \\
		a12m310, a12m220S, a12m220, a12m220L		 	& $35$ & $4.47$ && $ 25 $ & $3.78$ && $ 10 $ & $ 1.0 $  \\
		a09m310, a09m220 			& $70$ & $6.32$ && $ 50 $ & $5.35$ && $ 10 $ & $ 1.0 $  \\ \hline \hline 
	\end{tabular}

	\caption{\label{tab:MILC_smearingParams}Gaussian smearing parameters for the quark fields appearing in the interpolating operators.}
\end{table}

% **********

\subsubsection{\label{sec:corrmat}Correlation matrices}

As discussed in the introduction, a major technical advance of this work compared to previous lattice QCD investigations of $\bar b \bar b u d$ and $\bar b \bar b u s$ tetraquarks is that we use the two types of operators, local operators and scattering operators, both at the source and at the sink of the corresponding correlation functions. Thus, we compute square correlation matrices
\begin{equation}
\label{eq:defCorrelationMatrix} C_{jk} (t) = \Big\langle \op_j(t) \op_k\dagg(0) \Big\rangle
\end{equation}
with operators (\ref{eq:op_BastB_total_zero}) to (\ref{eq:op_BastBast_sep_zero}) for the $\bar b \bar b u d$ system and (\ref{eq:op_BBsast_total_zero}) to (\ref{eq:op_BastBsast_sep_zero}) for the $\bar b \bar b u s$ system, i.e., $5 \times 5$ or $7 \times 7$ symmetric correlation matrices, respectively. This is a major improvement compared to our previous papers \cite{Leskovec:2019ioa,Meinel:2022lzo}, where scattering operators were only used at the sink, and to similar work from other groups \cite{Francis:2016hui,Junnarkar:2018twb,Mohanta:2020eed,Hudspith:2023loy}, where only local operators were employed.

We consider the use of scattering operators to be important, even though here we are mainly interested in the ground states of the $\bar b \bar b u d$ and $\bar b \bar b u s$ sectors, which correspond to bound states in the infinite-volume limit. The reason why scattering operators might be helpful is that the ground state and the low-lying scattering states in the finite volume are expected to have close-by energies and also similar structures. The latter is also reflected in the operator definitions. For example, $\op_{[B B^\ast](0)}$ is contained in $\op_{B(0) B^\ast(0)}$ (considering only terms with $\xb = \yb$ on the right-hand side of Eq.\ (\ref{eq:op_BastB_sep_zero}) corresponds to the right-hand side of Eq.\ (\ref{eq:op_BastB_total_zero})). When extracting energy levels without using both local and scattering operators one might obtain an incorrect estimate of the lowest energy somewhere between the true ground state and one of the low-lying scattering states, resulting in underestimated tetraquark binding energies. In our previous works \cite{Leskovec:2019ioa,Meinel:2022lzo} we presented numerical evidence that the ground-state energies extracted from multi-exponential matrix fits are too high when using only local operators (see, in particular, Fig.~3 in Ref.~\cite{Meinel:2022lzo}).

More importantly, if one is interested in reliably determining the \emph{excited-state} finite-volume energy levels, the inclusion of scattering operators at both source and sink becomes essential. The excited-state energy levels can be used to determine the energy dependence of the meson-meson scattering amplitudes using L\"uscher's method \cite{Luscher:1990ux,Briceno:2017max} and rigorously search for shallow bound states or resonances, as we have done recently for the $\bar b \bar c u d$ systems with $I(J^P) = 0(0^+)$ and $I(J^P) = 0(1^+)$ \cite{Alexandrou:2023cqg}. Another interesting system to which these techniques could be applied is $\bar b \bar b u d$ with $I(J^P) = 0(1^-)$, as discussed in Section~\ref{sec:intro}.

% **********

\subsubsection{\label{SEC499}Quark propagators and computation of the correlation matrices}

We computed the correlation matrix elements with local operators at the source using Gaussian smeared point-to-all propagators, and the correlation matrix elements with scattering operators at the source using Gaussian smeared stochastic timeslice-to-all propagators (see e.g.\ Ref.\ \cite{Abdel-Rehim:2017dok} for a comprehensive discussion of these techniques in the context of four-quark states). For the latter, we use spin and color dilution and employ the one-end-trick to reduce the statistical noise. Because the correlation matrix is symmetric, the matrix elements that combine a local and a scattering operator can be computed either way, with point-to-all propagators (placing the local operator at the source) or stochastic timeslice-to-all propagators (placing the scattering operator at the source). We did both and found statistical uncertainties of similar magnitude. To increase the statistical precision we averaged both results on each gauge-link configuration.

To generate the necessary sources for the point-to-all-propagators, we selected 30 points randomly on each gauge link configuration (an exception is ensemble a15m310 with a rather small number of lattice sites, where we selected only 10 points). Moreover, for the stochastic timeslice-to-all propagators we use 4 equally separated timeslices with a randomly chosen global offset in $t$ direction (an exception is ensemble a15m310, where we use only 1 timeslice). On each timeslice we generate 5 independent random $\mathbb{Z}_2 \times \mathbb{Z}_2$ sources. When using stochastic timeslice-to-all propagators, i.e.\ two times the one-end-trick, it is essential that the two stochastic sources are independent. Since we have used 5 sources per timeslice, there are $5 \times 4 = 20$ possible combinations. We consider and average over all these combinations to further reduce the statistical noise. 

To determine statistical uncertainties, we use the jackknife method throughout this work.

% ********************
% ********************
% ********************

\subsection{$ B $ and $ B_s $ mesons}

To determine the binding energies of the QCD-stable $ \bar{b} \bar{b} u d $  and $ \bar{b} \bar{b} u s $ tetraquarks, we need to compare the ground-state energies of these four-quark systems to the respective lowest meson-meson thresholds. To this end, we also computed the energies of the pseudoscalar and vector $ B $ and $ B_s $ mesons using exactly the same lattice parameters. The corresponding interpolating operators are
\begin{align}
\label{eq:Binterpolator}  &\op_{B(\pb)}	= \sum_{\xb} \bar{b}(\xb) \gamma_5 u(\xb) \textrm{e}^{i\xb\cdot \pb} , \\
\label{eq:BastInterpolator} & \op_{B^\ast(\pb)} = \sum_{\xb} \bar{b}(\xb) \gamma_j u(\xb) \textrm{e}^{i\xb\cdot \pb} , \\
\label{eq:Bsinterpolator} & \op_{B_s(\pb)} = \sum_{\xb} \bar{b}(\xb) \gamma_5 s(\xb) \textrm{e}^{i\xb\cdot \pb} , \\
\label{eq:BsastInterpolator} & \op_{B_s^\ast(\pb)} = \sum_{\xb} \bar{b}(\xb) \gamma_j s(\xb) \textrm{e}^{i\xb\cdot \pb} .
\end{align}
We allow for non-vanishing momenta $\pb = 2 \pi \mathbf{n} / L$ with $L = N_s a$, $\mathbf{n} \in \mathbb{Z}^3$, which are necessary to compute the kinetic masses of the $B$ and the $B_s$ mesons. These kinetic masses are required for scattering analyses, which we carried out to estimate finite-volume effects on binding energies as outlined in \cref{sec:FVeffects}. For the computation of the correlation functions we use the same Gaussian smeared point-to-all propagators discussed in \cref{SEC499}.

% ********************
% ********************
% ********************
% ********************
% ********************

\FloatBarrier

\section{\label{SEC332}Energies and kinetic masses of pseudoscalar and vector $ B $ and $ B_s $ mesons}

We determined the ground-state energies of pseudoscalar and vector $ B $ and $ B_s $ mesons via correlated $\chi^2$-minimizing single-exponential fits to the correlation functions of the interpolating operators (\ref{eq:Binterpolator}) to (\ref{eq:BsastInterpolator}) with $\pb = 0$. We consider a number of fit ranges $t_\text{min} \leq t \leq t_\text{max}$ by varying $ 7 \leq t_\text{min} / a \leq 9 $ and $ 17 \leq t_\text{max} / a \leq 20$. The lower bound $t_\text{min} = 7 a$ corresponds to the temporal separation for which the corresponding effective energies start to be consistent with a plateau. We generate final results by weighted averaging following a method used by the FLAG Collaboration \cite{FlavourLatticeAveragingGroup:2019iem} (a brief summary of the method can also be found in Appendix~B of Ref.\ \cite{Meinel:2022lzo}).

\begin{figure}[htb]
		\centering
		\includegraphics[width=0.49\textwidth]{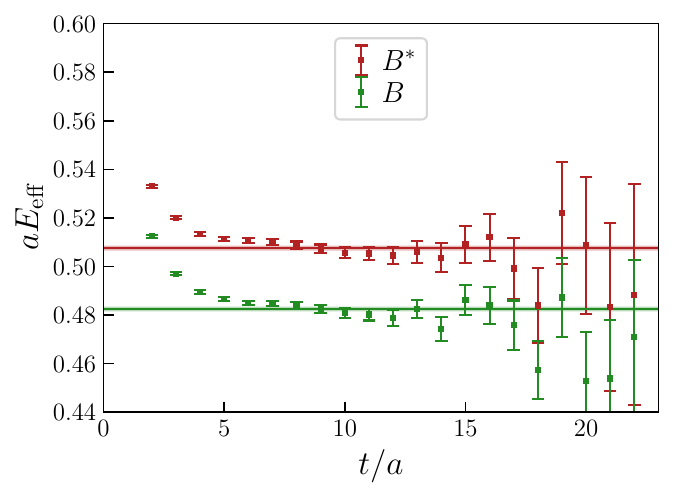}
		\includegraphics[width=0.49\textwidth]{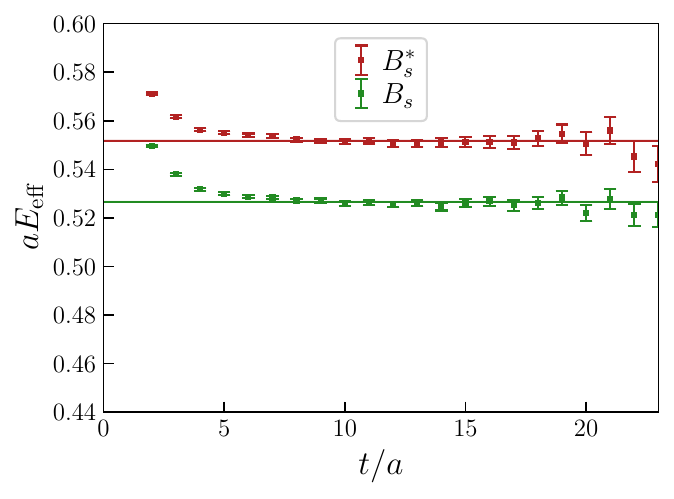}
		\caption{\label{fig:effm_plots_mesons_MILC}Effective energies of pseudoscalar and vector $ B $ and $ B_s $ mesons computed on ensemble a12m220L. Horizontal lines represent the corresponding final energies listed in Table~\ref{tab:MILC_mesonEnergies}.}
	\end{figure}

The results for the $B$, $B^\ast$, $B_s$ and $B_s^\ast$ meson energy levels for each of our seven ensembles are listed in \cref{tab:MILC_mesonEnergies}. To exemplify the quality of our numerical data, we show in Fig.~\ref{fig:effm_plots_mesons_MILC} plots of effective energies $a E_\text{eff}(t) = \ln(C(t) / C(t+a))$ for ensemble a12m220L together with the corresponding final energies from Table~\ref{tab:MILC_mesonEnergies}.

To check and to confirm the stability of our results with respect to $t_\text{min}$, we carried out additional fits with $t_\text{min}/a = 10$ and $t_\text{min}/a = 11$ and included them in our weighted averages for $aE_B(0)$ and $aE_{B^*}(0)$. The corresponding results are very similar to those listed in Table~\ref{tab:MILC_mesonEnergies} with differences much smaller than the statistical errors.

\begin{table}[htb]
	\centering
	\begin{tabular}{c|ccc|ccc}
		\hline\hline
		ensemble &    $aE_B(0)$     &  $aE_{B^*}(0) $  & $E_{B^\ast} - E_B$ [MeV] & $ aE_{B_s}(0)$ & $aE_{B_s^*}(0) $ & $E_{B_s^\ast} - E_{B_s}$ [MeV] \\ \hline
		a15m310  & $0.56575(62)\vz$ & $0.59583(72)\vz$ & $39.3(0.3)$ &   $0.61547(29)$ & $0.64452(34)$ & $38.0(0.1)$ \\ \hline
		a12m310  & $0.48902(88)\vz$ & $0.51561(99)\vz$ & $43.5(0.8)$ &   $0.53050(43)$ & $0.55641(51)$ & $42.4(0.4)$ \\
		a12m220S &  $0.48269(113)$  &  $0.50820(126)$  & $41.9(1.0)$ &   $0.52653(44)$ & $0.55215(50)$ & $42.1(0.4)$ \\
		a12m220  &  $0.48197(118)$  &  $0.50710(130)$  & $41.9(0.9)$ &   $0.52655(61)$ & $0.55203(51)$ & $42.5(0.4)$ \\
		a12m220L & $0.48246(98)\vz$ &  $0.50754(103)$  & $41.6(0.9)$ &   $0.52657(41)$ & $0.55180(45)$ & $41.9(0.4)$ \\ \hline
		a09m310  & $0.39766(53)\vz$ & $0.41713(59)\vz$ & $43.3(0.6)$ &   $0.42922(29)$ & $0.44894(33)$ & $43.8(0.3)$ \\
		a09m220  & $0.38835(77)\vz$ &  $0.40690(102)$  & $42.0(0.9)$ &   $0.42346(35)$ & $0.44239(42)$ & $42.8(0.5)$ \\ \hline\hline
	\end{tabular}
	\caption{\label{tab:MILC_mesonEnergies}Energies of pseudoscalar and vector $ B $ and $ B_s $ mesons and their differences.}
\end{table}

Note that, due to the use of NRQCD, the energies listed in Table~\ref{tab:MILC_mesonEnergies} and elsewhere contain overall negative shifts proportional to $n_b$, the number of $b$ quarks present in the corresponding states. At tree-level, this shift amounts to $-n_b m_b$, where $m_b$ is the $b$-quark mass. When considering energy differences between four-quark states and meson-meson thresholds with the same number of $b$ quarks, as done, for example, in \cref{sec:results}, these energy shifts cancel. Further quantities that can directly be compared to their experimental counterparts are the energy differences between the vector and pseudoscalar $B$ and $B_s$ meson energies. Experimentally, these differences are $E_{B^\ast} - E_B = 45.21(21) \, \text{MeV}$ and $E_{B_s^\ast} - E_{B_s} = 48.5^{+1.8}_{-1.5} \, \text{MeV}$ \cite{ParticleDataGroup:2022pth}. Our corresponding lattice results are listed in \cref{tab:MILC_mesonEnergies} and are found to be approximately 10\% smaller. This discrepancy is likely due to the missing one-loop corrections to the matching coefficient $c_4$ of the term $- g / (2 m_b)\, \psi^\dag\boldsymbol{\sigma}\cdot\mathbf{B}\psi $ in the NRQCD action.

\begin{figure}[htb]
			\centering
			\includegraphics[width=0.49\textwidth]{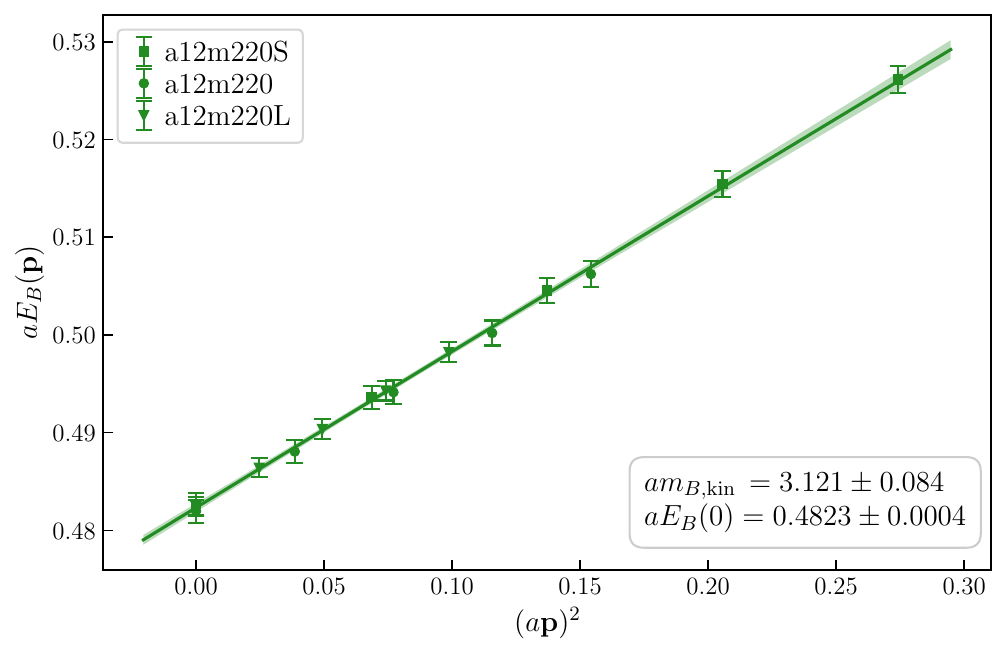}
			\caption{\label{fig:DispRel_B_a12m220_all}The $B$-meson energies for ensembles a12m220S, a12m220, and a12m220L as function of $(a \pb)^2$, together with the fit using Eq.\ (\protect\ref{EQN855}).}
		\end{figure}

We also determined the kinetic meson masses $m_{M,\,{\rm kin}}$ for $M \in \{ B, B^\ast, B_s , B_s^\ast \}$, in which the energy shifts are not present, and which reflect the energy-momentum dispersion relation. The kinetic meson masses are important for scattering analyses, as discussed in \cref{sec:FVeffects}. To determine them, we additionally computed the meson energies for several non-vanishing momenta, using the interpolating operators (\ref{eq:Binterpolator}) to (\ref{eq:BsastInterpolator}) with $\pb = 2 \pi \mathbf{n} / L$ with $\mathbf{n} \in \{ (1,0,0), (1,1,0), (1,1,1), (2,0,0) \}$. Then we extracted $m_{M,\,{\rm kin}}$ by fitting the right-hand side of
\begin{align}
\label{EQN855} E_M(\pb) - E_M(0) = \sqrt{m_{M,\,{\rm kin}}^2 + \pb^2} - m_{M,\,{\rm kin}}
\end{align}
for each $M \in \{ B, B^\ast, B_s , B_s^\ast \}$ and each ensemble to the corresponding lattice results for the four energy differences $E_M(\pb) - E_M(0)$. An exception are the three ensembles a12m220S, a12m220 and a12m220L, which only differ in the volume, and where we determined a common kinetic mass by performing a single fit for each meson to the corresponding twelve energy differences. The results for the kinetic masses are collected in Table~\ref{tab:MILC_kinMasses}. Moreover, in Fig.~\ref{fig:DispRel_B_a12m220_all} we present an example plot demonstrating that the energy-momentum relation (\ref{EQN855}) is fully consistent with the data.

	\begin{table}[h]
		\centering
		\begin{tabular}{c|cc|cc}
			\hline\hline
			ensemble & $am_{B,\textrm{kin}}$ & $a m_{B^*,\textrm{kin}} $ & $am_{B_s,\textrm{kin}}$ & $am_{B_s^*,\textrm{kin}} $ \\ \hline
			a15m310   &    $4.015(47)\vz$     &      $4.017(55)\vz$       &     $4.100(24)\vz$      &       $4.100(30)\vz$       \\ \hline
			a12m310   &    $3.224(96)\vz$     &       $3.238(108)$        &     $3.264(49)\vz$      &       $3.266(57)\vz$       \\
			a12m220S, a12m220, a12m220L &    $3.121(84)\vz$     &      $3.091(89)\vz$       &     $3.233(37)\vz$      &       $3.204(40)\vz$       \\ \hline
			a09m310   &    $2.381(55)\vz$     &      $2.382(60)\vz$       &     $2.426(31)\vz$      &       $2.427(35)\vz$       \\
			a09m220   &     $2.342(161)$      &       $2.341(214)$        &     $2.397(79)\vz$      &       $2.399(94)\vz$       \\ \hline\hline
		\end{tabular}
		\caption{\label{tab:MILC_kinMasses}Kinetic masses of the pseudoscalar and vector $ B $ and $ B_s $ mesons.}
	\end{table}

% ********************
% ********************
% ********************
% ********************
% ********************

\FloatBarrier

\section{\label{sec:results}Energies of antiheavy-antiheavy-light-light four-quark systems}

To determine the low-lying energy levels for the $ \bbud $ and $ \bbus $ four-quark systems from the  $ 5 \times 5 $ and $ 7 \times 7 $ correlation matrices $C_{j k}(t)$ discussed in Section~\ref{sec:corrmat}, we solve the standard generalized eigenvalue problems (GEVPs)
\begin{align}
\label{EQN016} \sum_k C_{j k}(t) v_{k,n}(t,t_0) = \lambda_n(t,t_0) \sum_k C_{j k}(t_0) v_{k,n}(t,t_0),
\end{align}
where the indices $j$ and $k$ are operator indices and the indices $n$ enumerate the resulting eigenvalues and eigenvectors and are, thus, related to the indices of the low-lying energy eigenstates. The choice of $t_0$ is made individually for each system and each ensemble. Our strategy was to choose the largest value of $t_0$ for which statistical uncertainties on the resulting eigenvalues and eigenvectors did not significantly increase (see Table~\ref{TAB002}). The energy levels $E_n$ are then determined by carrying out a correlated $\chi^2$-minimizing fit of an exponential function to each resulting eigenvalue $\lambda_n(t,t_0)$. Below, we also discuss the components $v_{j,n}(t,t_0)$, where we always normalize the eigenvectors according to $\sum_j |v_{j,n}(t,t_0)|^2 = 1$.

	\begin{table}[htb]
		\centering
		\begin{tabular}{cccc} \hline \hline
			ensemble & $ t_0/a $ for $ \bbud $ & & $ t_0/a $ for $ \bbus $ \\ \hline 
			a15m310 & 2 && 2 \\
			a12m310, a12m220L, a12m220, a12m220S 		& 3	&& 3 \\ 
			a09m310, a09m220 		& 4 && 6	 \\ \hline \hline
		\end{tabular}
		\caption{\label{TAB002}The $t_0$ values used for the GEVPs.}
	\end{table}

% ********************
% ********************
% ********************

\subsection{\label{SEC834}$ \bbud $ energy levels}

\begin{figure}[htb]
	\centering
	\includegraphics[width=0.7\textwidth]{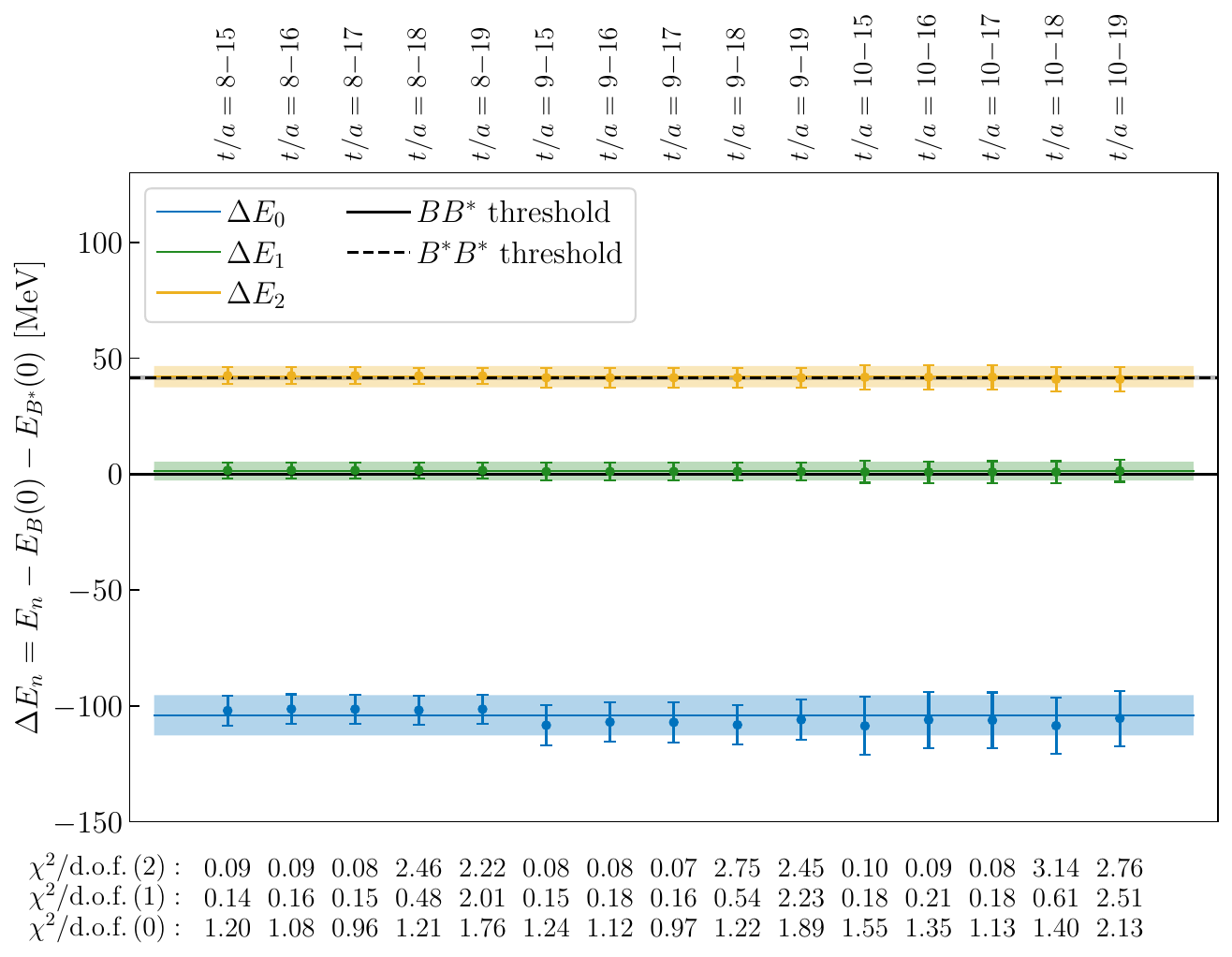}
	\caption{\label{fig:fitResults_bbud_a12m220L}Fit ranges and fit results for the extraction of energy levels for the $ \bbud $ system for ensemble a12m220L. The $B B^\ast$ and $B^\ast B^\ast$ thresholds correspond to the $B$ and $B^\ast$ energies obtained on that ensemble (see Table~\protect\ref{tab:MILC_mesonEnergies}).}
\end{figure}

\begin{figure}[htb]
 \includegraphics[width=0.7\linewidth]{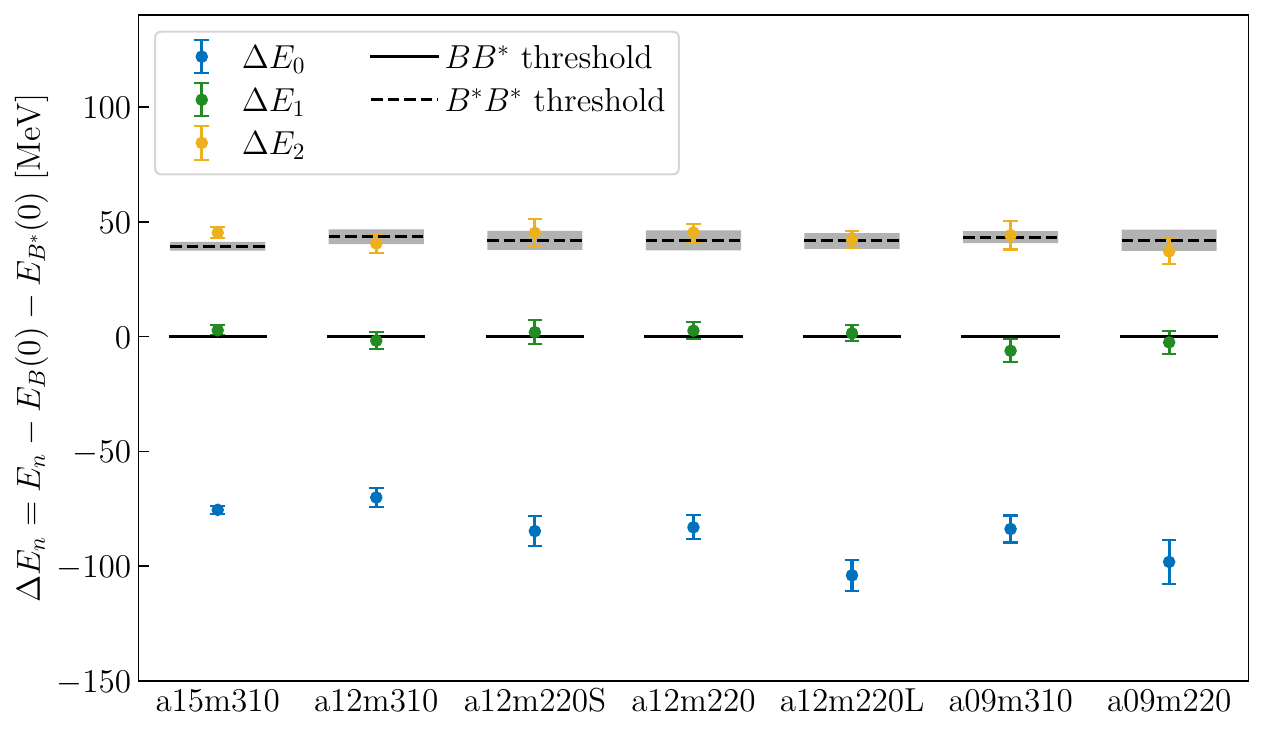}
 	\caption{\label{fig:MILC_allResults_bbud}The three lowest finite-volume energy levels for the $ \bbud $ system with $ I(J^P)=0(1^+) $ obtained from our operator basis on each ensemble relative to the $ BB^* $ threshold, $\Delta E_n = E_n - E_B(0) - E_{B^\ast}(0)$. The black horizontal lines correspond to the $ BB^* $ (solid) and $ B^*B^* $ (dashed) thresholds.}
\end{figure}

As done in our determination of $B$ and $B_s$ meson energies in \cref{SEC332}, we again consider multiple fit ranges $t_\text{min} \leq t \leq t_\text{max}$ by varying $t_\text{min}$ and $t_\text{max}$. For ensemble a12m220L, the fits are summarized in a graphical way in Fig.~\ref{fig:fitResults_bbud_a12m220L}, including the corresponding fit ranges and resulting energy levels. Analogous plots for the other ensembles are collected in Fig.~\ref{fig:fitResults_bbud_appendix} in the appendix. As before, the final results are then generated by weighted averaging over all fits according to the FLAG method and are shown in Table~\ref{tab:finiteVolumeResults_bbud} and Fig.~\ref{fig:MILC_allResults_bbud}, where $\Delta E_n = E_n - E_B(0) - E_{B^\ast}(0)$.

Again we confirmed the stability of our results with respect to $t_\text{min}$ by carrying out additional analyses:
\begin{itemize}
\item[(i)] We carried out further fits with two larger $t_\text{min}$ values, $t_\text{min}/a = 11$ and $t_\text{min}/a = 12$, and included them in the weighted average for the ground state energy $E_0$ of the $ \bbud $ system. The shift caused by these larger $t_\text{min}$ values is tiny, for ensemble a12m220L $\approx -0.1 \, \text{MeV}$.

\item[(ii)] Same as (i), but we removed all fits with $t_\text{min}/a = 7$. The weighted average leads to a slightly lower result for $E_0$, $\approx -3.0 \, \text{MeV}$ below the the value represented by the blue line in Fig.~\ref{fig:fitResults_bbud_a12m220L}.

\item[(iii)] Same as (i), but we removed all fits with $t_\text{min}/a = 7$ and $t_\text{min}/a = 8$. The weighted average leads to a slightly lower result for $E_0$, $\approx -2.4 \, \text{MeV}$ below the the value represented by the blue line in Fig.~\ref{fig:fitResults_bbud_a12m220L}.

\item[(iv)] We use the same fits as in (i) but use the recently proposed technique of Bayesian model averaging (see Ref.\ \cite{Jay:2020jkz}) instead of the FLAG method. This technique was partly developed and is particularly suited as an alternative to manual selection of fit ranges. This time we find an upward shift of $\approx +0.2 \, \text{MeV}$.
\end{itemize}
As expected, these four results are fully consistent within statistical errors with the result shown in Fig.~\ref{fig:fitResults_bbud_a12m220L} and used for further analysis later in this work. In units of the statistical error the energy shifts are $-0.01$, $-0.22$, $-0.12$ and $+0.02$ for (i), (ii), (iii) and (iv) respectively. We conclude that our analysis method provides very stable and, thus, reliable results, which we attribute partly to the $\chi^2$-dependent FLAG weighting, which favors fits with good $\chi^2$, while those with a certain tension are suppressed.

\begin{table}[htb]
	\centering
	\begin{tabular}{cccc}
		\hline\hline
		& $ \Delta E_0 $ [MeV] & $ \Delta E_1 $ [MeV] & $ \Delta E_2 $ [MeV] \\ \hline
		a15m310  & $ \vz -75.5(1.8)\vz$  &    $\vp2.7(2.2)$     &     $45.2(2.5)$      \\
		a12m310  & $ \vz -70.1(4.2)\vz$  &     $-1.8(3.8)$      &     $40.5(4.2)$      \\
		a12m220S & $ \vz -84.7(6.5)\vz$  &    $\vp1.9(5.3)$     &     $45.1(5.9)$      \\
		a12m220  & $ \vz -83.1(5.2)\vz$  &    $\vp2.5(3.7)$     &     $45.2(3.9)$      \\
		a12m220L &    $-104.0(6.8)$      &    $\vp1.4(3.5)$     &     $42.1(3.7)$      \\
		a09m310  & $ \vz -83.8(5.9)\vz$  &     $-6.2(5.0)$      &     $44.1(6.2)$      \\
		a09m220  & $ \vz -98.1(9.6)\vz$  &     $-2.5(5.0)$      &     $37.1(5.5)$      \\ \hline\hline
	\end{tabular}
	\caption{\label{tab:finiteVolumeResults_bbud}Finite-volume energy levels for the $ \bbud $ system with respect to the $B B^\ast$ threshold, $\Delta E_n = E_n - E_B(0) - E_{B^\ast}(0)$.}
\end{table}

The ground-state energy is found to be significantly below the $B B^\ast$ threshold on all ensembles, $\Delta E_0 \approx (-105 \ldots \mbox{-70})$ MeV.
An energy level in that region is expected from and consistent with previous lattice QCD investigations of this $ \bbud $ system \cite{Francis:2016hui,Junnarkar:2018twb,Leskovec:2019ioa,Mohanta:2020eed,Hudspith:2023loy,Aoki:2023nzp}. It signals the existence of a QCD-stable $ \bbud $ tetraquark. The energies of the first and second excitations are within statistical uncertainties consistent with the $B B^\ast$ and $B^\ast B^\ast$ thresholds. This indicates that, with our basis of interpolating operators, we are able to not only resolve the tetraquark state, but also these two low-lying scattering states.

As expected, with the operator basis used here [Eqs.~(\ref{eq:op_BastB_total_zero}) to (\ref{eq:op_BastBast_sep_zero})], we are not able to resolve scattering states with non-vanishing relative momenta. In particular, this is the case for $B B^\ast$ states in which the two mesons have opposite minimal non-vanishing momenta $p_\text{min} = (2 \pi) / L$, where $L = N_s a$ is the ensemble-dependent spatial extent of the lattice. The estimated energy of such states with respect to the $B B^\ast$ threshold is
\begin{align}
\label{EQN422} \Delta E_{B B^\ast,p=p_\text{min}} = \Big(m_B^2 + p_\text{min}^2\Big)^{1/2} + \Big(m_{B^\ast}^2 + p_\text{min}^2\Big)^{1/2} - m_B - m_{B^\ast} ,
\end{align}
where $m_B$ and $m_{B^\ast}$ are the full meson masses, which can be taken to be the kinetic masses from Table~\ref{tab:MILC_kinMasses}. For example, for ensembles a12m220 and a12m220L, $\Delta E_{B B^\ast,p=p_\text{min}} \approx 20 \, \text{MeV}$ and $\Delta E_{B B^\ast,p=p_\text{min}} \approx 13 \, \text{MeV}$, respectively. With exception of ensemble a15m310, where $\Delta E_{B B^\ast,p=p_\text{min}} \approx 50 \, \text{MeV}$, we have $\Delta E_{B B^\ast,p=p_\text{min}} < m_{B^\ast} - m_B \approx 45 \, \text{MeV}$. Thus, the true second excited state is expected significantly below the $B^\ast B^\ast$ threshold, which is not reflected by the numerically obtained energy differences $\Delta E_2$ (see Table~\ref{tab:finiteVolumeResults_bbud}). The overlaps of our operators with this state are likely very small. To resolve this state and higher states of this type, we would need to include scattering operators with non-vanishing relative momenta.

\begin{figure}[htb]
	\centering
	\includegraphics[width=0.49\textwidth]{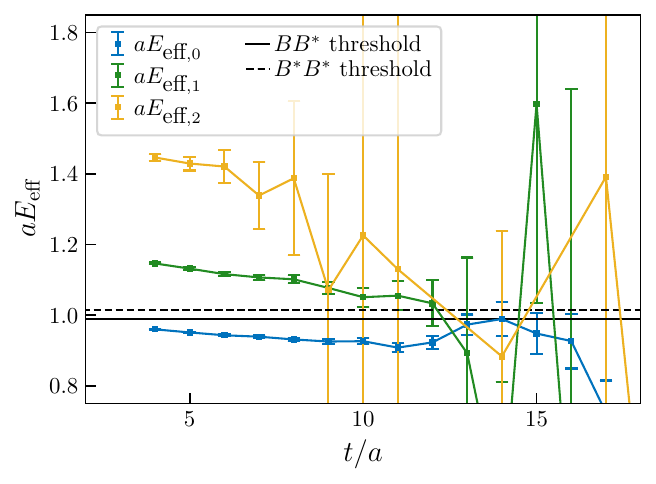}
	\includegraphics[width=0.49\textwidth]{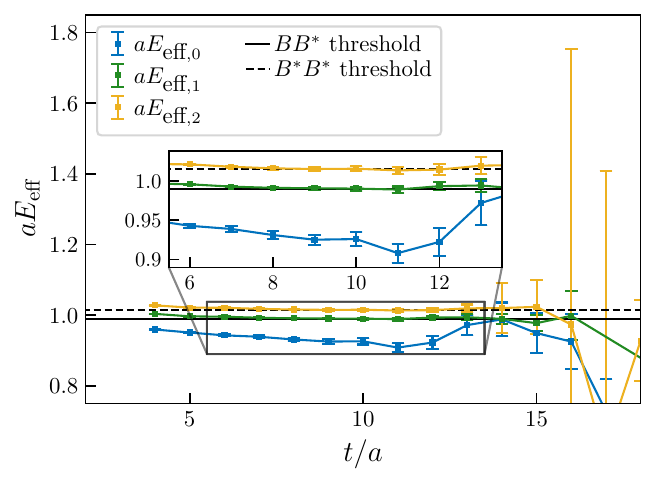}
	\caption{\label{fig:effm_bbud}Effective energies $E_{\text{eff},n}(t)$, $n = 0,1,2$ for the $ \bbud $ system (ensemble a12m220L). \textbf{Left}: $3 \times 3$ correlation matrix with only local interpolating operators [Eqs.~(\ref{eq:op_BastB_total_zero}) to (\ref{eq:op_Dd_total_zero})]. \textbf{Right}: $5 \times 5$ correlation matrix including also scattering interpolating operators [Eqs.~(\ref{eq:op_BastB_sep_zero}) and (\ref{eq:op_BastBast_sep_zero})]. The inset in the right plot shows a magnified view on the effective masses indicating that they are consistent with plateaus within statistical errors for $t \geq t_\text{min}$ for our chosen $t_\text{min}/a = 8, 9, 10$.}
\end{figure}

To illustrate the importance of scattering operators, we compare in Fig.~\ref{fig:effm_bbud} effective energies obtained by solving GEVPs for a $3 \times 3$ correlation matrix with only local operators [Eqs.~(\ref{eq:op_BastB_total_zero}) to (\ref{eq:op_Dd_total_zero})] (left plot) and the full $5 \times 5$ correlation matrix, which also contains the scattering operators [Eqs.~(\ref{eq:op_BastB_sep_zero}) and (\ref{eq:op_BastBast_sep_zero})] (right plot). While the effective masses $E_{\text{eff},1}(t)$ and $E_{\text{eff},2}(t)$, defined as $E_{\text{eff},n}(t) = (1/a) \ln(\lambda_n(t,t_0) / \lambda_n(t+a,t_0))$, quickly approach the $B B^\ast$ and $B^\ast B^\ast$ thresholds for the full $5 \times 5$ matrix, this is not the case for the $3 \times 3$ matrix. Thus, it is practically impossible to extract physically meaningful energy levels $E_n$ for $n \geq 1$ from the $3 \times 3$ matrix. The conclusion is that it is imperative to include scattering operators if energy levels of scattering states are needed, e.g., for a finite-volume scattering analysis as discussed in \cref{sec:FVeffects}. However, the results for the effective energy $E_{\text{eff},0}(t)$ are essentially the same with and without scattering operators. Thus, if one is only interested in determining the mass of the QCD-stable $\bbud$ tetraquark in a finite volume, local operators may be sufficient. This finding is surprising, given our experience in Refs.~\cite{Leskovec:2019ioa,Meinel:2022lzo}, where we compared multi-exponential matrix fits with and without scattering operators at the sink and found some impact on the fit results for $E_0$. In Refs.~\cite{Leskovec:2019ioa,Meinel:2022lzo}, we were unable to use the GEVP for the non-square correlation matrices with scattering operators included at the sink only.

\begin{figure}[htb]
	\centering
	\includegraphics[width=0.7\textwidth]{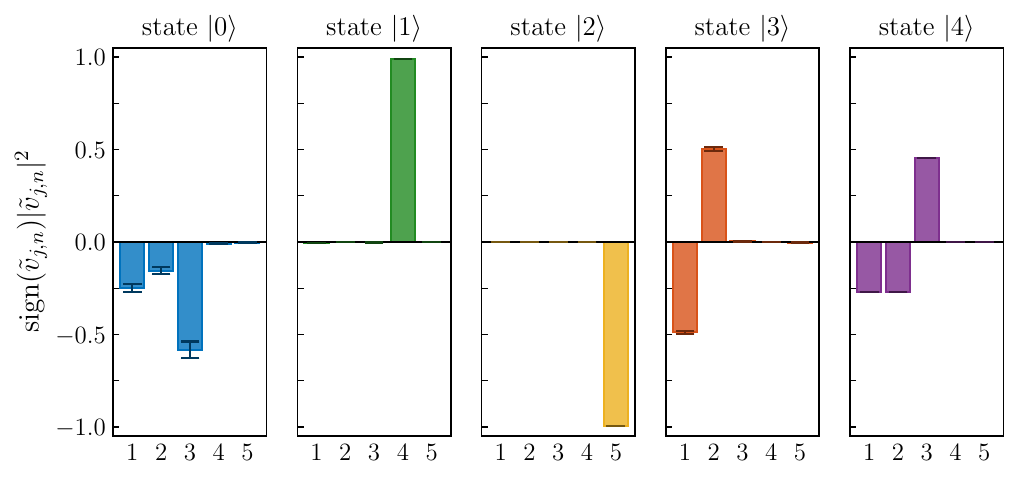}
	\caption{\label{fig:EV_bbud_a12m220L}Signed squared eigenvector components, $\text{sign}(\tilde{v}_{j,n}) |\tilde{v}_{j,n}|^2$, from ensemble a12m220L.}
\end{figure}

The eigenvectors $v_{j,n}(t,t_0)$ obtained by solving GEVPs are fairly independent of $t$ for larger $t$ and provide information about the composition of the low-lying energy eigenstates associated with the extracted energy levels $E_n$. In Fig.~\ref{fig:EV_bbud_a12m220L} we show the signed squared eigenvector components, $\text{sign}(\tilde{v}_{j,n}) |\tilde{v}_{j,n}|^2$, for the full $5 \times 5$ correlation matrix from ensemble a12m220L, where the quantities $\tilde{v}_{j,n}$ are obtained by fitting constants to $v_{j,n}(t,t_0)$ in the range $t_\text{min} \leq t \leq t_\text{max}$ (we use multiple fits with different ranges, $ 6 \leq t_\text{min}/a \leq 9 $ and $ 12 \leq t_\text{max}/a \leq 14 $ in this case, and present the FLAG-method averages over these fits). When ignoring the signs, these quantities add up to 1 and can be interpreted as the relative importance of each interpolating operator $\mathcal{O}_j$ when approximating the energy eigenstate $|n\rangle$ as a sum over trial states $\mathcal{O}_j | n \rangle$ \footnote{Such an interpretation requires a similar normalization of all trial states $\mathcal{O}_j | n \rangle$. One possibility to implement this is to include appropriate volume factors, as present in the definitions of our interpolating operators. A common alternative is to replace $C_{jk}(t)$ by $ C_{jk}(t) / \sqrt{C_{jj}(t=a) C_{kk}(t=a)}$, which corresponds to a normalization of the trial states at $t=a$, i.e.\ $\langle \Omega | \mathcal{O}_j(t=a) \mathcal{O}_k^\dagger(t=0) | \Omega \rangle = 1$. Throughout this paper we follow this latter strategy. Note that the eigenvalues $\lambda_n(t,t_0)$ and, consequently, the extracted energy levels $E_n$ are unaffected by such a normalization.}. The signs are of particular interest in the $ \bbus $ case (see \cref{SEC533}), where they expose the approximate light flavor symmetry of $u$ and $s$ quarks.

The results support our above conclusions based on the extracted energy levels. The ground state $| 0 \rangle$, which is a QCD-stable tetraquark, is excited almost exclusively by the three local operators [Eqs.~(\ref{eq:op_BastB_total_zero}) to (\ref{eq:op_Dd_total_zero})], while the scattering operators [Eqs.~(\ref{eq:op_BastB_sep_zero}) and (\ref{eq:op_BastBast_sep_zero})] are essentially irrelevant. The relative contributions of the local operators to the QCD-stable tetraquark are, moreover, consistent with previous lattice QCD investigations using static potentials and the Born-Oppenheimer approximation. In particular, the $B B^\ast$ component is slightly larger than, but of similar magnitude as the $B^\ast B^\ast$ component, which is in agreement with Fig.~3 of Ref.\ \cite{Bicudo:2016ooe}, and the meson-meson components are of similar importance as the diquark-antidiquark component, which is in line with the main result from Ref.\ \cite{Bicudo:2021qxj}. The first and second excitation $| 1 \rangle$ and $| 2 \rangle$, on the other hand, are clearly $B B^\ast$ and $B^\ast B^\ast$ scattering states, as already indicated by the consistency of their energy levels with the corresponding thresholds.

% ********************
% ********************
% ********************

\subsection{\label{SEC533}$ \bbus $ energy levels}

\begin{figure}[htb]
	\centering
	\includegraphics[width=0.7\textwidth]{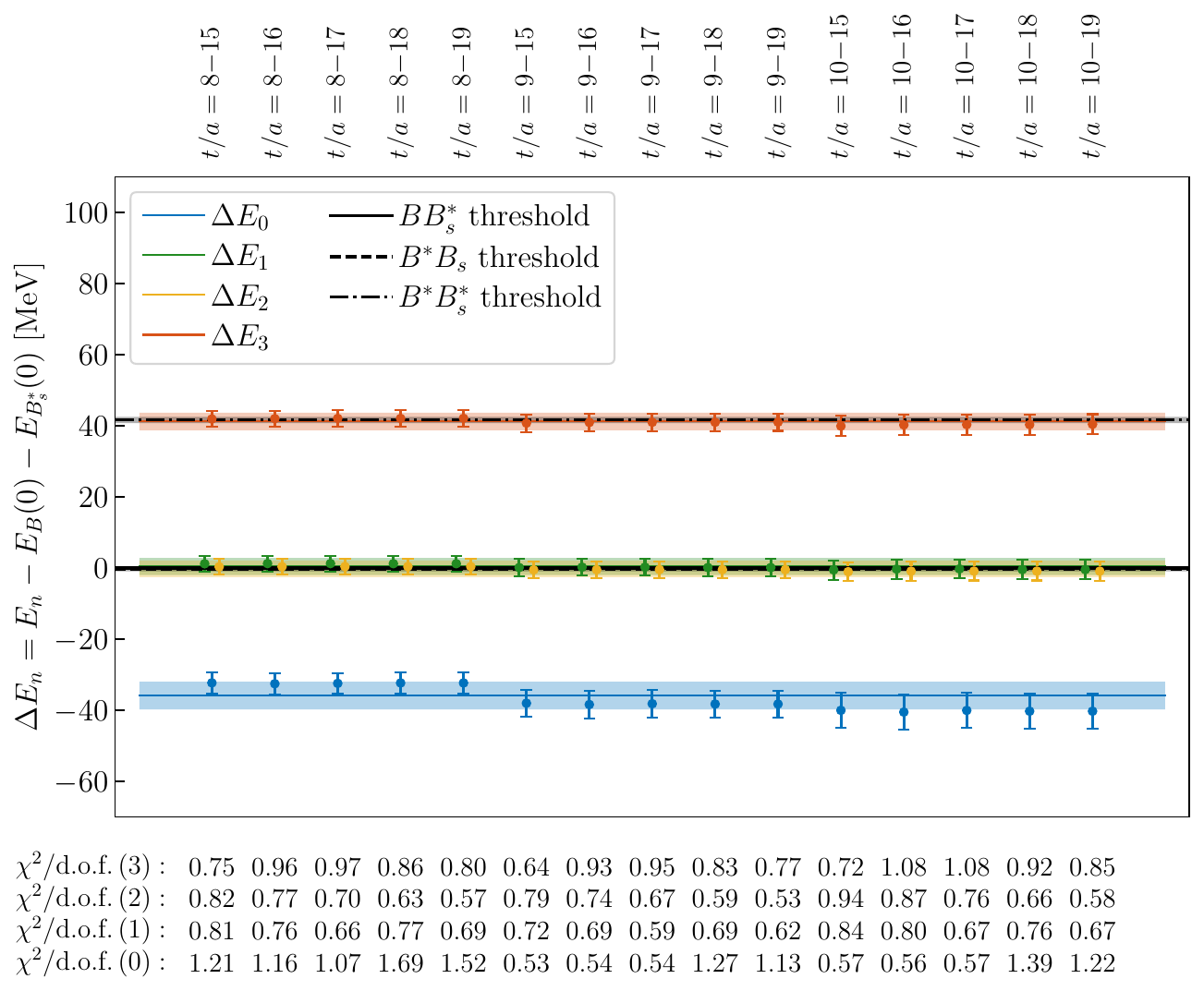}
	\caption{\label{fig:fitResults_bbus_a12m220L}Fit ranges and fit results for the extraction of energy levels for the $ \bbus $ system for ensemble a12m220L. The $B B_s^\ast$, $B_s B^\ast$ and $B^\ast B_s^\ast$ thresholds correspond to the $B_{(s)}$ and $B_{(s)}^\ast$ energies obtained on that ensemble (see Table~\protect\ref{tab:MILC_mesonEnergies}).}
\end{figure}

\begin{figure}
 \includegraphics[width=0.7\linewidth]{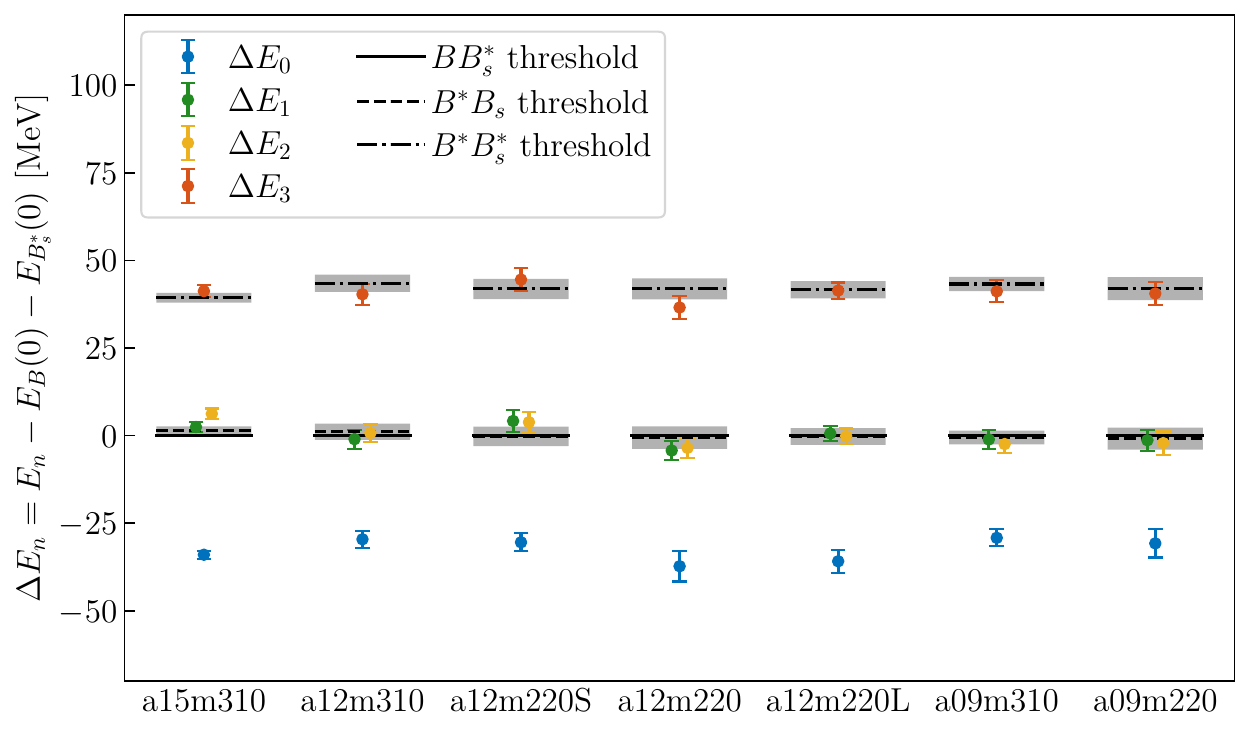}
	\caption{\label{fig:MILC_allResults_bbus}The four lowest finite-volume energy levels for the $ \bbus $ system with $ J^P=1^+ $ obtained from our operator basis on each ensemble relative to the $ BB_s^* $ threshold, $\Delta E_n = E_n - E_B(0) - E_{B_s^\ast}(0)$. The black horizontal lines correspond to the $ BB_s^* $ (solid), $ B^*B_s $ (dashed) and $ B^*B_s^* $ (dashed-dotted) thresholds.}
\end{figure}

For the $ \bbus $ system we proceed in the same way as for the $ \bbud $ system in \cref{SEC834}. For ensemble a12m220L, the fits are summarized in Fig.~\ref{fig:fitResults_bbus_a12m220L}; the plots for the other ensembles are collected in Fig.~\ref{fig:fitResults_bbus_appendix} in the appendix. The final results for $\Delta E_n = E_n - E_B(0) - E_{B_s^\ast}(0)$ from the FLAG-method weighted averages are shown in Table \ref{tab:finiteVolumeResults_bbus} and Fig.~\ref{fig:MILC_allResults_bbus}.

As in the $ \bbud $ case, we checked and confirmed the stabilty of these results by carrying out the same analyses (i), (ii), (iii) and (iv) discussed in \cref{SEC834}. We find fully consistent ground state energies $E_0$ only slightly shifted (by $\approx -0.4 \, \text{MeV} ,  -3.1 \, \text{MeV} ,  -3.6 \, \text{MeV} ,  -2.1 \, \text{MeV}$) with respect to the value represented by the blue line in Fig.~\ref{fig:fitResults_bbus_a12m220L}. Again these shifts are well within the statistical errors, in units of the corresponding statistical error $-0.08$, $-0.56$, $-0.55$ and $-0.40$ for (i), (ii), (iii) and (iv) respectively.

The ground-state energy is found to be  significantly below the $B B_s^\ast$ threshold, $\Delta E_0 \approx (-35 \ldots -30) \text{MeV}$, indicating the existence of a QCD-stable $ \bbus $ tetraquark, consistent with previous lattice-QCD investigations of this system \cite{Francis:2016hui,Junnarkar:2018twb,Meinel:2022lzo,Hudspith:2023loy}. We obtain precise results also for three additional states with energies consistent with the $B B_s^\ast$, $B_s B^\ast$ and $B^\ast B_s^\ast$ thresholds. However, as already noted for the $ \bbud $ case, we are unable to resolve scattering states with non-vanishing back-to-back momenta; for example, on ensembles a12m220 and a12m220L one has $\Delta E_{B B_s^\ast,p=p_\text{min}} \approx 20 \, \text{MeV}$ and $\Delta E_{B B_s^\ast,p=p_\text{min}} \approx 13 \, \text{MeV}$, respectively. Scattering operators with non-vanishing back-to-back momenta would be needed to resolve these states.

\begin{table}
	\centering
	\begin{tabular}{ccccc}
		\hline\hline
		& $ \Delta E_0 $ [MeV] & $ \Delta E_1 $ [MeV] & $ \Delta E_2 $ [MeV] & $ \Delta E_3 $ [MeV] \\ \hline
		a15m310  &     $-34.0(1.2)$     &    $\vp2.4(1.4)$     &    $\vp6.2(1.5)$     &     $41.2(1.7)$      \\
		a12m310  &     $-29.6(2.5)$     &     $-1.0(2.7)$      &    $\vp0.7(2.4)$     &     $40.3(2.9)$      \\
		a12m220S &     $-30.4(2.6)$     &    $\vp4.2(3.0)$     &    $\vp3.8(2.8)$     &     $44.5(3.2)$      \\
		a12m220  &     $-37.3(4.3)$     &     $-4.2(2.8)$      &     $-3.5(2.9)$      &     $36.6(3.2)$      \\
		a12m220L &     $-35.8(3.2)$     &    $\vp0.6(2.3)$     &     $-0.1(2.2)$      &     $41.4(2.3)$      \\
		a09m310  &     $-29.1(2.5)$     &     $-1.1(2.7)$      &     $-2.4(2.6)$      &     $41.2(3.2)$      \\
		a09m220  &     $-30.7(4.0)$     &     $-1.3(3.0)$      &     $-2.1(3.3)$      &     $40.5(3.3)$      \\ \hline\hline
	\end{tabular}
	\caption{\label{tab:finiteVolumeResults_bbus}Finite-volume energy levels for the $ \bbus $ system with respect to the $B B_s^\ast$ threshold, $\Delta E_n = E_n - E_B(0) - E_{B_s^\ast}(0)$.}
\end{table}

\begin{figure}
	\centering
	\includegraphics[width=0.49\textwidth]{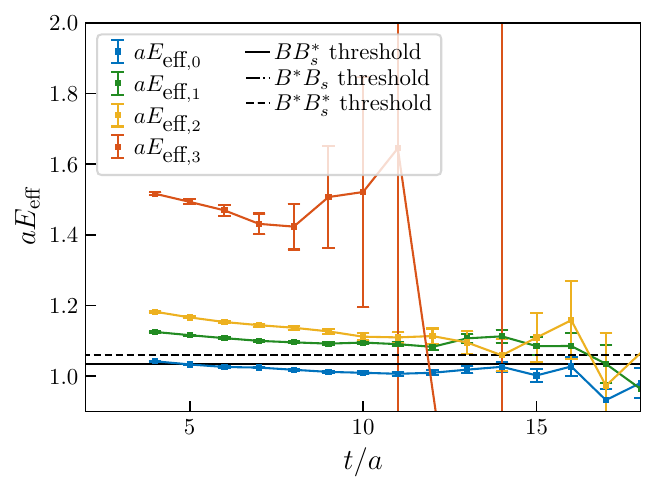}
	\includegraphics[width=0.49\textwidth]{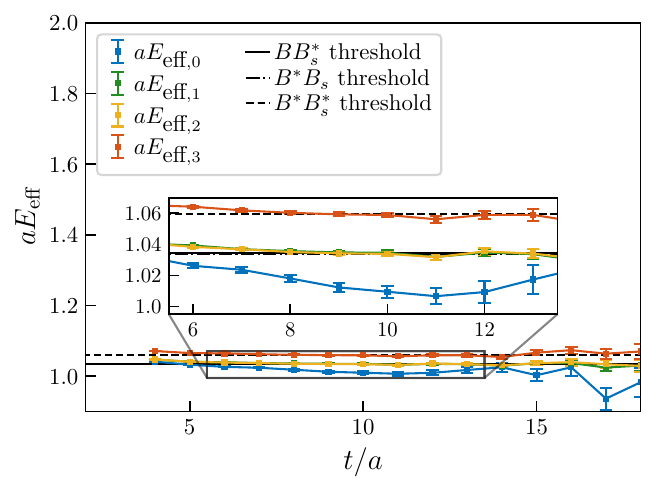}
	\caption{\label{fig:effm_4x4and7x7}Effective energies $E_{\text{eff},n}(t)$, $n = 0,1,2,3$ for $ \bbus $ (ensemble a12m220L). \textbf{Left}: $4 \times 4$ correlation matrix with only local interpolating operators [Eqs.~(\ref{eq:op_BBsast_total_zero}) to (\ref{eq:op_Dds_total_zero})]. \textbf{Right}: $7 \times 7$ correlation matrix with also scattering interpolating operators [Eqs.~(\ref{eq:op_BBsast_sep_zero}) to (\ref{eq:op_BastBsast_sep_zero})]. The inset in the right plot shows a magnified view on the effective masses indicating that they are consistent with plateaus within statistical errors for $t \geq t_\text{min}$ for our chosen $t_\text{min}/a = 8, 9, 10$.}
\end{figure}

\begin{figure}
	\centering
	\includegraphics[width=0.85\textwidth]{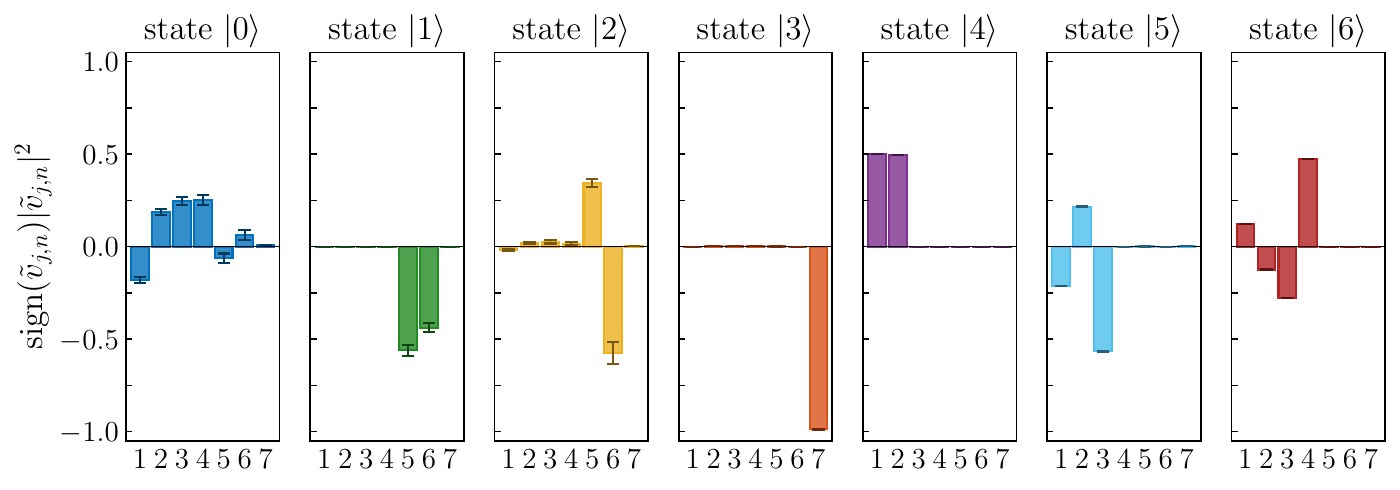}
	\caption{\label{fig:EV_bbus_a12m220L}Signed squared eigenvector components, $\text{sign}(\tilde{v}_{j,n}) |\tilde{v}_{j,n}|^2$, from ensemble a12m220L.}
\end{figure}

Again, we illustrate the importance of the scattering operators by comparing in Fig.~\ref{fig:effm_4x4and7x7} effective energies obtained by solving GEVPs for a $4 \times 4$ correlation matrix with only local operators [Eqs.~(\ref{eq:op_BBsast_total_zero}) to (\ref{eq:op_Dds_total_zero})] (left plot) and the full $7 \times 7$ correlation matrix, which also contains the scattering operators [Eqs.~(\ref{eq:op_BBsast_sep_zero}) to (\ref{eq:op_BastBsast_sep_zero})] (right plot). The overall picture and conclusions are the same as for the $ \bbud $ system. To determine the mass of the QCD-stable $\bbus$ tetraquark, local operators appear to be sufficient, but it is imperative to include scattering operators if energy levels of scattering states are needed. The fact that the results for $E_{\text{eff},0}(t)$ are essentially the same with and without scattering operators is even more surprising here than for the $ \bbud $ system, given the weaker binding.

In Fig.~\ref{fig:EV_bbus_a12m220L} we show the signed squared eigenvector components $\text{sign}(\tilde{v}_{j,n}) |\tilde{v}_{j,n}|^2$ for the full $7 \times 7$ correlation matrix from ensemble a12m220L (here, we use the FLAG-method averages over fits with $ 5 \leq t_\text{min}/a \leq 9 $ and $ 11 \leq t_\text{max}/a \leq 14 $). The eigenvector components support and complement our above conclusions based on the extracted energy levels, and exhibit the approximate SU(3) flavor symmetry. The ground state $| 0 \rangle$, which is a QCD-stable tetraquark, is excited mostly by the four local operators. It is $u s$ flavor antisymmetric, which shows that the $ \bbus $ tetraquark has a flavor structure analogous the $I = 0$ $ \bbud $ tetraquark discussed in the previous section. The first and second excitations are mixtures of $B B_s^\ast$ and $B_s B^\ast$ scattering states, where $| 1 \rangle$ is mostly $u s$ flavor symmetric, i.e.\ similar to $I = 1$, while $| 2 \rangle$ is mostly $u s$ flavor antisymmetric, i.e.\ similar to $I = 0$. The third excitation $| 3 \rangle$ is clearly a $B^\ast B_s^\ast$ scattering state, as already suggested by its energy.

\FloatBarrier

% ********************
% ********************
% ********************
% ********************
% ********************

\section{\label{sec:FVeffects}Finite-volume effects}

Some of the previous lattice-QCD studies of $\bar{b}\bar{b}ud$ and $\bar{c}\bar{c}ud$ systems \cite{Leskovec:2019ioa,Padmanath:2022cvl,Chen:2022vpo} have employed L\"uscher's method \cite{Luscher:1990ux,Rummukainen:1995vs,Kim:2005gf,Christ:2005gi,Hansen:2012tf,Leskovec:2012gb,Gockeler:2012yj,Briceno:2014oea,Briceno:2017tce,Briceno:2017max} to relate the finite-volume energy levels with infinite-volume $B$-$B^*$ and $D$-$D^*$ scattering amplitudes. The masses of the infinite-volume bound states or virtual bound states were then obtained as the pole locations in these amplitudes. Such an analysis has not yet been performed for the $\bar{b}\bar{b}us$ system, which is more challenging due to the coupling of the $B_s$-$B^*$ and $B$-$B_s^*$ channels with nearly identical thresholds \cite{Meinel:2022lzo}. The improvements made in the present work in extracting the finite-volume spectra (compared to Refs.~\cite{Leskovec:2019ioa,Meinel:2022lzo}) in principle enable more advanced scattering-amplitude analyses, including an analysis of coupled-channel $B_s$-$B^*$, $B$-$B_s^*$ scattering.

The studies presented in Refs.~\cite{Leskovec:2019ioa,Padmanath:2022cvl,Chen:2022vpo} used L\"uscher's method both above and below threshold, and used effective-range expansions (EREs) to parametrize the energy dependence of the scattering amplitude. It was recently pointed out that the $D$-$D^*$ scattering amplitude has a left-hand cut starting close to threshold due to pion exchange \cite{Du:2023hlu}, which determines the radius of convergence of the ERE to be smaller than the regions studied in the aforementioned references. The same problem is also present in $B$-$B^*$ scattering and, due to kaon exchange, in $B_s$-$B^*$, $B$-$B_s^*$ scattering. Moreover, the two-body L\"uscher quantization condition itself becomes invalid on the left-hand cut ~\cite{Raposo:2023nex,Raposo:2023oru,Dawid:2023jrj,Dawid:2023lcq,Meng:2023bmz,Hansen:2024ffk,Bubna:2024izx}.

For $B$-$B^*$ scattering, the left-hand cut due to single-pion exchange starts at the invariant-mass-squared
\begin{equation}
 s_{\rm cut} = s_{\rm th} - m_\pi^2 + (m_{B^*}-m_B)^2,
\end{equation}
where $s_{\rm th}=(m_{B^*}+m_B)^2$. For physical pion mass, one finds $\sqrt{s_{\rm cut}}-\sqrt{s_{\rm th}}\approx -1\:{\rm MeV}$, while at $m_\pi=300\:{\rm MeV}$, $\sqrt{s_{\rm cut}}-\sqrt{s_{\rm th}}\approx -4\:{\rm MeV}$. Similarly, for $B_s$-$B^*$, $B$-$B_s^*$ scattering at physical kaon mass, the cut due to single-kaon exchange starts at  $\sqrt{s_{\rm cut}}-\sqrt{s_{\rm th}}\approx -11\:{\rm MeV}$. The finite-volume ground-state energies we obtained in this work (Section~\ref{sec:results}) and in Refs.~\cite{Leskovec:2019ioa,Meinel:2022lzo} for $\bar{b}\bar{b}ud$ and $\bar{b}\bar{b}us$ are well below these values.

New approaches to describing the finite-volume energy levels in the region effected by the left-hand cut were recently proposed in Refs.~\cite{Raposo:2023nex,Raposo:2023oru,Dawid:2023jrj,Dawid:2023lcq,Meng:2023bmz,Hansen:2024ffk,Bubna:2024izx}; these involve modifications to the two-body L\"uscher formalism, the use of the three-body L\"uscher formalism, or chiral effective theory. However, the implementation of these approaches is beyond the scope of the present work.

In the remainder of this paper, we will focus on the ground-state energies and assume that finite-volume effects are negligible compared to our statistical uncertainties. Before we became aware of the left-hand cut problem, we did in fact perform extractions of $S$-wave scattering amplitudes from the energy levels presented here using the standard two-body L\"uscher method; we can use these results to get at least some idea of the possible size of finite-volume effects. For the 
$\bar{b}\bar{b}ud$ system, we used the lowest two energy levels on each ensemble and fitted a two-parameter model of the $K$ matrix \cite{Chew:1960iv}, similar to the effective-range expansion used in \cite{Leskovec:2019ioa}. From these fits, we found that the bound-state pole locations differ from the finite-volume ground-state energies by $\lesssim 0.1$ MeV, consistent with the expectation from Ref.~\cite{Leskovec:2019ioa}. For
$\bar{b}\bar{b}us$, we attempted coupled-channel fits to the lowest three energy levels with various parametrizations of the $2 \times 2$ $K$ matrix, including fits that combine the data from the a12m220S, a12m220, and a12m220L ensembles that differ only in the volume. However, most of these fits did not converge properly, and the only model that gave stable results was a diagonal form of the $K$ matrix with two $s$-independent parameters on the diagonal. For those ensembles for which we achieved a reasonable $\chi^2/{\rm d.o.f}$, this model predicted shifts between the finite-volume ground-state energies and the bound-state pole locations of $\lesssim 1$ MeV. Another way of estimating the finite-volume effects is directly comparing the finite-volume energies of the a12m220S, a12m220, and a12m220L ensembles listed in Table~\ref{tab:finiteVolumeResults_bbus}, which suggests that the effects for $\bar{b}\bar{b}us$ are not larger than about 5 MeV for the smallest volume. Further tests for finite-volume effects are discussed in Sec.~\ref{sec:extrap}.

\FloatBarrier
\section{\label{sec:extrap}Chiral and continuum extrapolations of the binding energies}
\FloatBarrier

To obtain estimates of the tetraquark binding energies at the physical point, we consider two different fits to the ground-state energies from the different ensembles: (i) neglecting lattice-spacing dependence but allowing for a linear dependence on $m_\pi^2$, as in Refs.~\cite{Leskovec:2019ioa,Meinel:2022lzo},
\begin{align}
\label{eq:chiralExtrap} \Delta E_0(m_\pi) = \Delta E_0(m_{\pi,\text{phys}}) + c (m_\pi^2 - m_{\pi,\text{phys}}^2),
\end{align}
and (ii), allowing for linear dependencies on both $m_\pi^2$ and $a^2$,
\begin{align}
\label{eq:chiralContinuumExtrap} \Delta E_0(m_\pi,a) = \Delta E_0(m_{\pi,\text{phys}},0) + c (m_\pi^2 - m_{\pi,\text{phys}}^2) + d\, a^2.
\end{align}
Here, we use $m_{\pi,\text{phys}} = 135 \, \textrm{MeV}$ for the physical pion mass in the isospin-symmetric limit. These models are expected to be approximate only. For the actions used here, gluon and sea-quark discretization errors are expected to start at $O(\alpha_s^2 a^2)$ and $O(\alpha_s a^2)$, respectively, but the Wilson-clover valence action is only tree-level-$O(a)$ improved and thus will have some remaining $O(\alpha_s a)$ errors. With the particular level of Symanzik improvement used here, the NRQCD action has both $O(\alpha_s^2 a)$ and $O(\alpha_s^2 a^2)$ errors, plus other systematic errors from missing higher-order relativistic and radiative corrections that we will estimate separately below.

\begin{figure}
	\centering
	\includegraphics[width=0.49\textwidth]{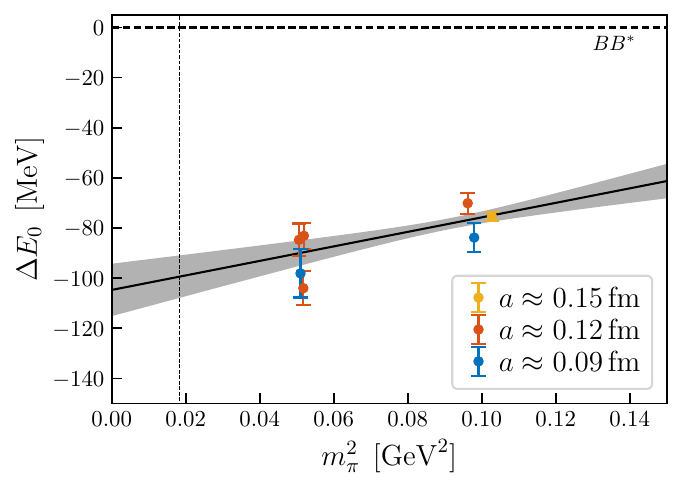}
	\includegraphics[width=0.49\textwidth]{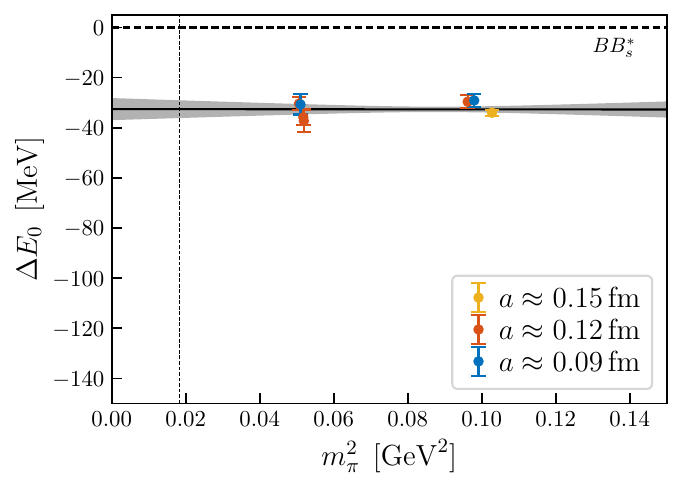}
	\caption{\label{fig:chiral_extrap}
		Chiral extrapolations of the energy differences $\Delta E_0 $ for the $ \bbud $ four-quark system with $ I(J^P)=0(1^+) $ (left) and the $ \bbus $ four-quark system with $ J^P=1^+ $(right), neglecting the lattice-spacing dependence. The lowest meson-meson threshold for each channel is indicated by the horizontal dashed lines, while the vertical dashed line represents the physical pion mass.}
\end{figure}

\begin{figure}
	\centering
	\includegraphics[width=0.49\textwidth]{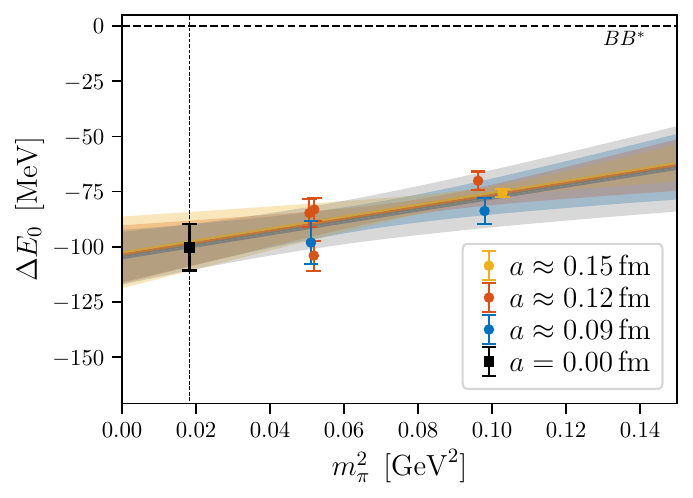}
	\includegraphics[width=0.49\textwidth]{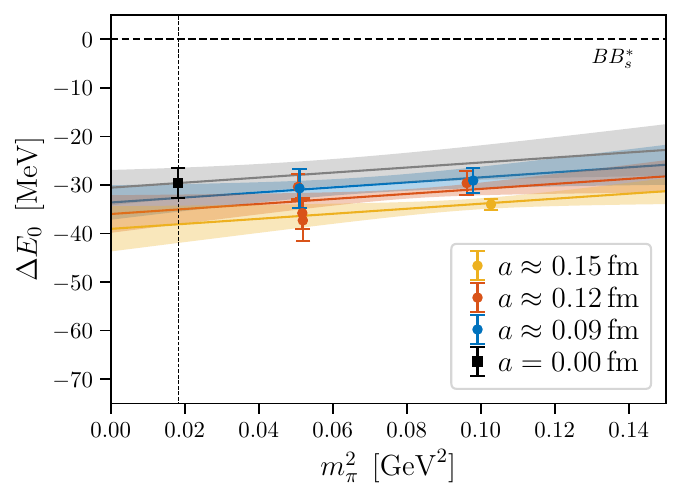}
	
	\includegraphics[width=0.49\textwidth]{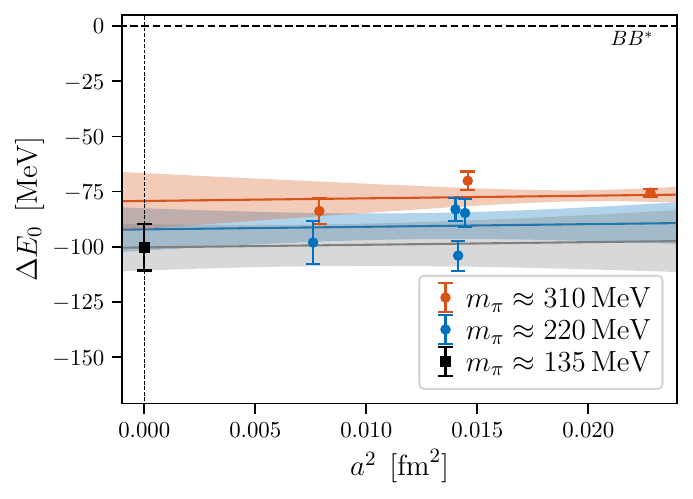}
	\includegraphics[width=0.49\textwidth]{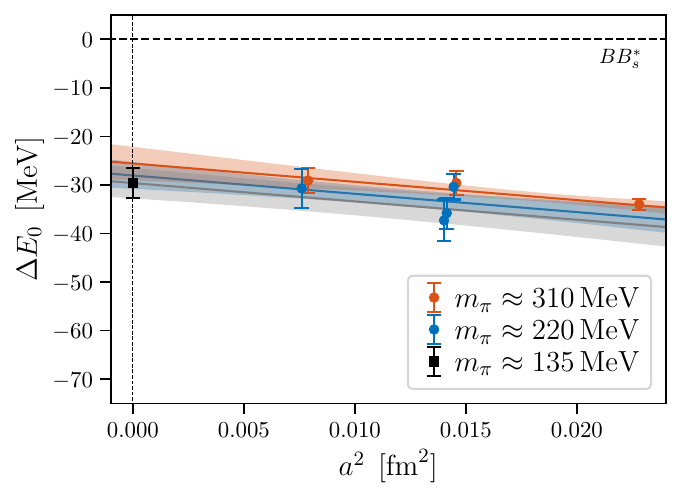}
	\caption{\label{fig:chiral_continuum_extrap}
		Combined chiral and continuum extrapolations of the energy differences $\Delta E_0 $ for the $ \bbud $ four-quark system with $ I(J^P)=0(1^+) $ (left) and the $ \bbus $ four-quark system with $ J^P=1^+ $(right). In the upper row, the $m_\pi$ dependence is shown for fixed lattice spacing $ a $, while the lower row shows the $a$ dependence for fixed pion mass $ m_\pi $. The lowest meson-meson threshold for each channel is indicated by the horizontal dashed lines, while the vertical dashed line represents either the physical pion mass or zero lattice spacing $ a $. The black square indicates the final result for the binding energy at the physical point.}
\end{figure}

Our fits using Eqs.~(\ref{eq:chiralExtrap}) or (\ref{eq:chiralContinuumExtrap}) are illustrated in Figs.~\ref{fig:chiral_extrap} and \ref{fig:chiral_continuum_extrap}, and the results for the fit parameters are given in Table \ref{tab:MILC_res_extrapolation_finite}. The coefficients $d$ that describe the lattice-spacing dependence turn out to be consistent with zero. For the $ \bbud $ system, there is a slight indication that the binding energy increases in magnitude as the pion mass is lowered, which becomes more statistically significant when not including the $a^2$ term. On the other hand, the $\bbus$ binding energy is essentially independent of the pion mass in the range considered here. Some of the fits have $ \chi^2/\textrm{d.o.f.}>1 $; in these cases, we scaled the uncertainties of the fit results by $\sqrt{\chi^2/\textrm{d.o.f.}}$. We find that the results for $\Delta E_0(m_{\pi,\text{phys}})$ and $\Delta E_0(m_{\pi,\text{phys}},0)$ are compatible within statistical uncertainties. We take the latter as our final estimates of the tetraquark binding energies.

\begin{table}
	\centering
	\begin{tabular}{ccccc}
		\hline\hline
		\multicolumn{5}{c}{Chiral extrapolation only}   \\ \hline
		& $ \Delta E_0(m_{\pi,\textrm{phys}}) $ [MeV]  				 & $ c $ [$10^{-4}$/MeV] &                               & $ \chi^2/\textrm{d.o.f.} $ \\ \hline
		$ \bbud $ & $  -99(8)$			& $ 2.9(1.1) $         	&                   & $ 2.34 $  \\
		$ \bbus $ & $  $$ -33(4) $ 	& $ 0.0(0.5) $ 	&                   & $ 1.57 $  \\ 
				\hline\hline
		\multicolumn{5}{c}{Chiral and continuum extrapolation}                                                   \\ \hline
		& $\Delta E_0(m_{\pi,\textrm{phys}},0) $ [MeV] & $ c $ [$10^{-4}$/MeV] & $ d $ [$10^{2}\times$MeV$^3$] & $ \chi^2/\textrm{d.o.f.} $ \\ \hline
		$ \bbud $ & $ -100(10) $ 		&    $ 2.7(1.6) $     	& $1.2 (6.2) $	& $ 2.90 $  \\
		$ \bbus $ & $ -30(3)\vz $ & $ 0.5(0.5) $	& $ 3.8(1.8) $     & $ 0.83 $  \\ \hline\hline
	\end{tabular}
	\caption{\label{tab:MILC_res_extrapolation_finite}Fit results for the chiral extrapolations using Eq.~(\ref{eq:chiralExtrap}) [upper table] and the combined chiral-continuum extrapolations using Eq.~(\ref{eq:chiralContinuumExtrap}) [lower table]. All uncertainties have been scaled by $\sqrt{\chi^2/\textrm{d.o.f.}}$. }
\end{table}

Given the large values of $ \chi^2/\textrm{d.o.f.} $ for the $\bbud$ system, we performed additional fits for this system to test for exponential finite-volume effects and higher-order discretization effects:
\begin{itemize}
 \item We excluded the data point with the smallest value of $m_\pi L$ (corresponding to the a12m220S ensemble) from the fits. We find that the $\chi^2/{\rm d.o.f.}$ values actually increase, because $\chi^2$ does not change much while d.o.f.~is reduced by 1. The results for $\Delta E_0$ change by about $-3$ MeV, which is negligible compared to the statistical uncertainty.
 \item We included overall factors of $(1+f\:e^{-m_\pi L})$ in Eqs.~(\ref{eq:chiralExtrap}) and (\ref{eq:chiralContinuumExtrap}), where $f$ is an additional fit parameter. The results for $f$ are consistent with zero at the 1.1$\sigma$ level, and $\chi^2/{\rm d.o.f.}$ again increases. The results for $\Delta E_0$ again change by about $-3$ MeV, which is negligible compared to the statistical uncertainty.
 \item For the fit using Eq.~(\ref{eq:chiralContinuumExtrap}), we excluded the data point with the coarsest lattice spacing (corresponding to the a15m310 ensemble). This improves  $\chi^2/{\rm d.o.f.}$ from 2.90 to 2.29 and leads to a shift in $\Delta E_0(m_{\pi,\text{phys}},0)$ from $-100(10)$ to $-128(21)$ MeV. We take the shift of $-28$ MeV as a systematic uncertainty due to possible higher-order discretization effects. For consistency, we perform the same steps also for the $\bbus$ system, where excluding the a15m310 data point leads to an increase in $\chi^2/{\rm d.o.f.}$ from 0.83 to 1.01 and a shift of $\Delta E_0(m_{\pi,\text{phys}},0)$ by $-3.3$ MeV.
\end{itemize}

The continuum extrapolations do not remove the systematic errors from  missing higher-order relativistic and radiative corrections in the lattice NRQCD action. To estimate these sources of systematic uncertainty, we follow the prescription of Ref.~\cite{Leskovec:2019ioa} with adjustments for the different choices of NRQCD matching coefficients used here.

The values of $\Delta E_0$ depend both on the heavy-light meson energies and on the energies of the four-quark systems. For the heavy-light mesons, the most significant systematic NRQCD uncertainties are expected to be as follows:
\begin{itemize}
	\item Four-quark operators are missing in the action. They appear only at order $ \alpha_s^2 $ in the QCD matching, but according to Ref.~\cite{Blok:1996iz} their effect is of order $ 3\,{\rm MeV} $.
	\item One-loop corrections to the matching coefficient $c_4$ of the operator $- g / (2 m_b)\, \psi^\dag\boldsymbol{\sigma}\cdot\mathbf{B}\psi $ are not included here. The associated uncertainties can be estimated as
	\begin{equation}
		\alpha_s \Lambda_{\textrm{QCD}}^2 / m_b \approx 6 \, {\rm MeV}, \label{eq:c4uncert}
	\end{equation} 
	where we use $ \alpha_s \approx0.3 $, $\Lambda_{\textrm{QCD}}\approx 300$ MeV, and $m_b\approx 4.5$ GeV.
	\item We also use only tree-level values for the coefficients $ c_2 $ and $ c_3 $ of the operators of order $ (\Lambda_{\textrm{QCD}}^2/ m_b^2) $. The uncertainties from missing higher-order contributions to these coefficients are given by
	\begin{equation}
		\alpha_s \Lambda_{\textrm{QCD}}^3 / m_b^2 \approx 0.4 \, {\rm MeV}.
	\end{equation}
\end{itemize}
As the matching coefficients $ c_1 $, $ c_5 $, and $ c_6 $ include order-$\alpha_s$ corrections, systematic uncertainties arising from the related operators are negligible.

For the four-quark system, the power counting is more complicated due to the presence of two bottom quarks. A reasonable estimate for the systematic uncertainties can be obtained by replacing the QCD scale $ \Lambda_{\textrm{QCD}} $ by the binding momentum $ |k_{\textrm{BS}}| $ (if larger than $ \Lambda_{\textrm{QCD}} $) \cite{Leskovec:2019ioa}. Our results for the binding energies correspond to $ |k_{\textrm{BS}}|\approx 730$ MeV for $\bbud$ and $ |k_{\textrm{BS}}|\approx 400$ MeV for $\bbus$, which leads to estimated systematic uncertainties of $\approx 36$ MeV for the $\bbud$ energy and $\approx 11$ MeV for the $\bbus$ energy. It is reasonable to expect some partial cancellation of systematic errors between the tetraquark and heavy-light meson energies, so these numbers could be taken as the total systematic uncertainties of $\Delta E_0$ \cite{Leskovec:2019ioa}. However, note that our previous calculations \cite{Leskovec:2019ioa,Meinel:2022lzo} predicted larger binding momenta  of $ |k_{\textrm{BS}}|\approx 800 $ MeV for $\bbud$ and $ |k_{\textrm{BS}}|\approx 680 $ MeV for $\bbus$, and the smaller values obtained here may, in principle, be due to the very systematic errors we are trying to estimate. Therefore, to be conservative, we will report asymmetric uncertainties in which we use the larger values of  $|k_{\textrm{BS}}|$ from Refs.~\cite{Leskovec:2019ioa,Meinel:2022lzo} to obtain the size of the error bars in the direction of increased binding. This prescription, combined in quadrature with the shifts seen when excluding a15m310, then yields our final results
\begin{eqnarray}
 \Delta E_0(m_{\pi,\text{phys}},0)[\bbud] &=& -100 \pm 10\:^{+36}_{-51}\:\:{\rm MeV}, \\
 \Delta E_0(m_{\pi,\text{phys}},0)[\bbus] &=& -30 \pm 3\:^{+11}_{-31}\:\:{\rm MeV}.
\end{eqnarray}
The NRQCD systematic uncertainties are dominated by the uncertainty from the missing radiative corrections to $c_4$, calculated using Eq.~(\ref{eq:c4uncert}) with $\Lambda_{\rm QCD}$ replaced by the $ |k_{\textrm{BS}}|$ values discussed above.

% ********************
% ********************
% ********************
% ********************
% ********************

\FloatBarrier

\section{\label{SEC596}Conclusions}

In this work, we have confirmed the existence and computed the binding energies of both the $\bar b \bar b u d$ tetraquark with quantum numbers $I(J^P) = 0(1^+)$ and the $\bar b \bar b u s$ tetraquark with quantum numbers $J^P = 1^+$ using lattice QCD. We have improved on existing similar lattice-QCD computations \cite{Francis:2016hui,Junnarkar:2018twb,Leskovec:2019ioa,Mohanta:2020eed,Meinel:2022lzo,Hudspith:2023loy,Aoki:2023nzp} by implementing, for the first time, a combination of local and scattering interpolating operators both at the source and at the sink of symmetric correlation matrices. We have demonstrated that this enables high-precision determinations of the energies of low-lying scattering states, and that it is practically impossible to determine these energies without scattering operators. Reliably extracting the complete low-lying finite-volume spectra is essential when using the L\"uscher method to determine energy-dependent scattering amplitudes, as we have done for the $\bar b \bar c u d$ systems with quantum numbers $I(J^P) = 0(0^+)$ and $I(J^P) = 0(1^+)$ in Ref.~\cite{Alexandrou:2023cqg}, based on the methods developed here.
On the other hand, we found that the inclusion of scattering operators had essentially no impact on the ground-state results obtained from the GEVP for the $\bar b \bar b u d$ and $\bar b \bar b u s$ systems. This finding was somewhat unexpected, given our experience from Refs.~\cite{Leskovec:2019ioa,Meinel:2022lzo}, in which we compared the results of multi-exponential matrix fits with and without scattering operators at the sink and found some impact on the fit results for $E_0$. When using the GEVP, local operators alone appear to be sufficient to determine the binding energies of the deeply bound $\bar b \bar b u d$ and $\bar b \bar b u s$ tetraquarks, but the inclusion of scattering operators provides reassurance and may be necessary to reliably extract the ground-state energies of other, less deeply bound systems.

\begin{figure}
	\centering
	\includegraphics[width=0.95\textwidth]{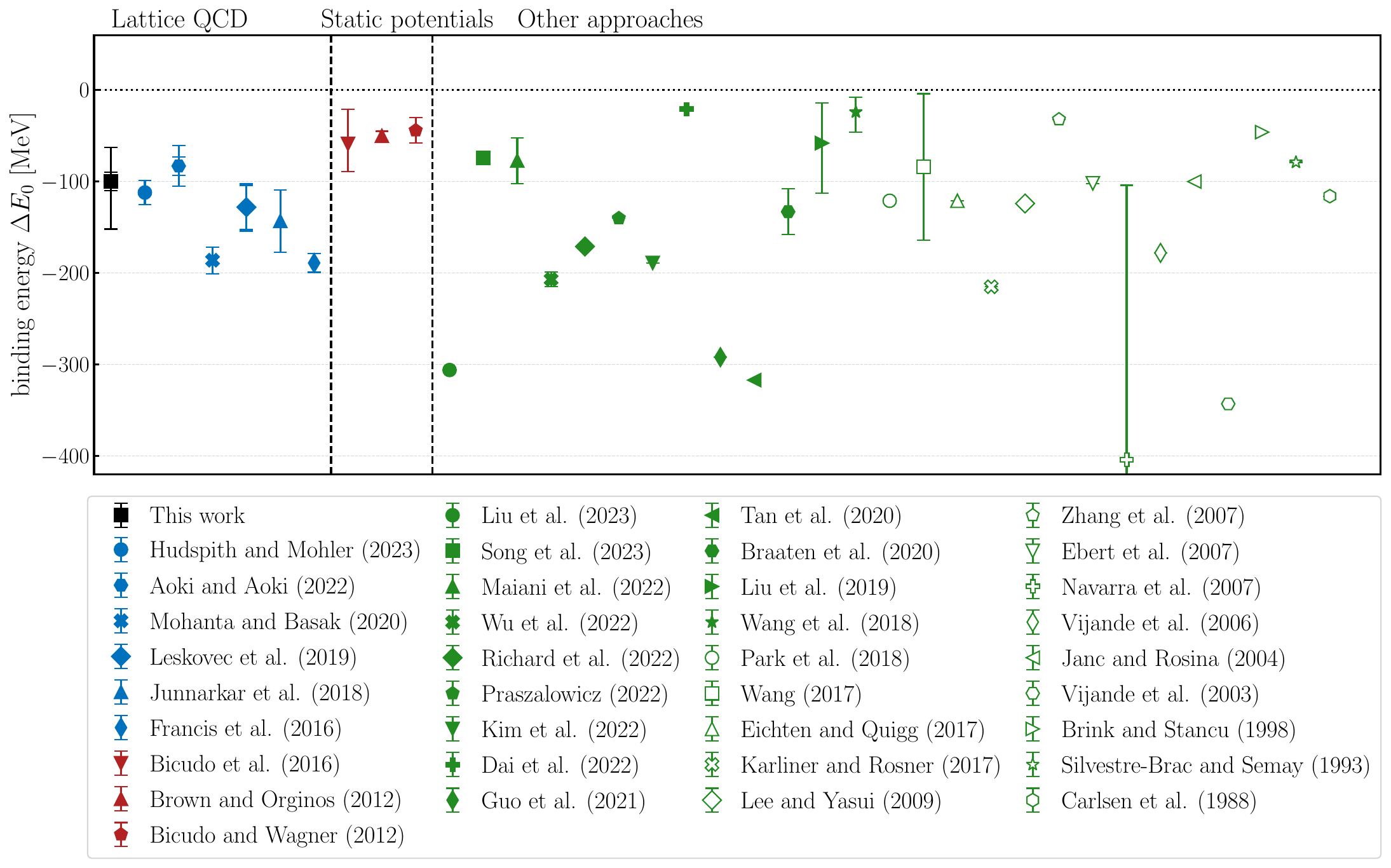} \\
  \vspace{0.4cm}	
	\includegraphics[width=0.8\textwidth]{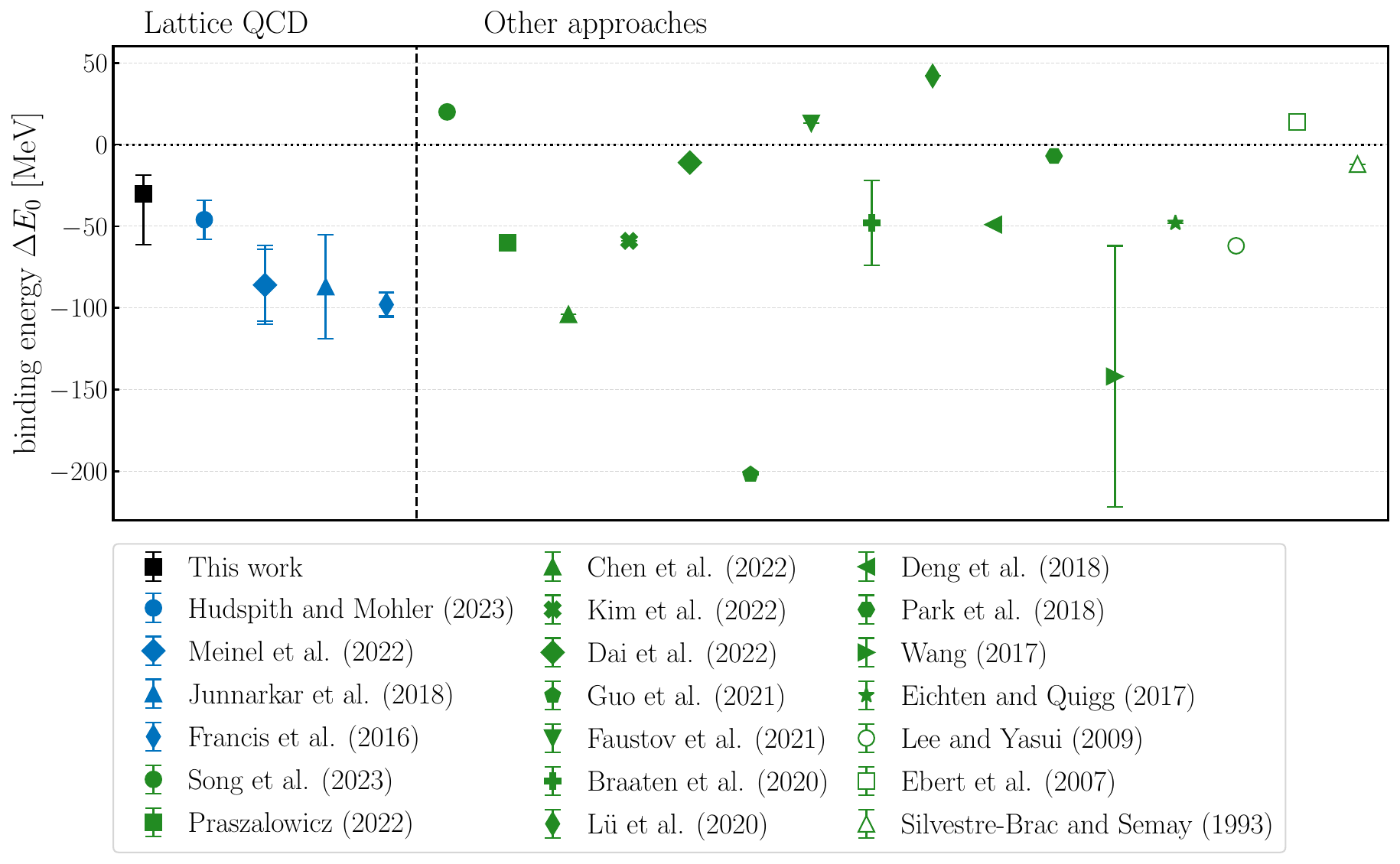}
	\caption{\label{FIG:compare}Comparison of results for the binding energy of antiheavy-antiheavy-light-light tetraquarks. \textbf{(upper plot)}: $\bar b \bar b u d$ tetraquark with $I(J^P) = 0(1^+)$ (black: this work; blue: previous full lattice-QCD simulations \cite{Francis:2016hui,Junnarkar:2018twb,Leskovec:2019ioa,Mohanta:2020eed,Hudspith:2023loy,Aoki:2023nzp}; red: lattice-QCD computations of static potentials in combination with the Born-Oppenheimer approximation \cite{Bicudo:2012qt,Brown:2012tm,Bicudo:2016ooe}; green: other approaches (quark models, phenomenological considerations, sum rules \cite{Carlson:1987hh,SilvestreBrac:1993zem,Brink:1998as,Vijande:2003ki,Janc:2004qn,Vijande:2006jf,Navarra:2007yw,Ebert:2007rn,Zhang:2007mu,Lee:2009rt,Karliner:2017qjm,Eichten:2017ffp,Wang:2017uld,Park:2018wjk,Wang:2018atz,Liu:2019stu,Tan:2020ldi,Guo:2021yws,Dai:2022ulk,Kim:2022mpa,Praszalowicz:2022sqx,Richard:2022fdc,Wu:2022gie,Maiani:2022qze,Song:2023izj,Liu:2023vrk}). \textbf{(lower plot)}: $\bar b \bar b u s$ tetraquark with $J^P = 1^+$ (black: this work; blue: previous full lattice-QCD simulations \cite{Francis:2016hui,Junnarkar:2018twb,Meinel:2022lzo,Hudspith:2023loy}; green: other approaches (quark models, phenomenological considerations, sum rules \cite{SilvestreBrac:1993zem,Ebert:2007rn,Lee:2009rt,Eichten:2017ffp,Wang:2017uld,Park:2018wjk,
Deng:2018kly,Lu:2020rog,Braaten:2020nwp,Faustov:2021hjs,Guo:2021yws,Dai:2022ulk,Kim:2022mpa,Chen:2022ros,Praszalowicz:2022sqx,Song:2023izj}).}
\end{figure}

Using chiral and continuum extrapolations of the results from several ensembles, we obtained binding energies of $-100 \pm 10\:^{+36}_{-51}\:\:{\rm MeV}$ for the $\bar b \bar b u d$ tetraquark with quantum numbers $I(J^P) = 0(1^+)$ and $-30 \pm 3\:^{+11}_{-31}\:\:{\rm MeV}$ for the $\bar b \bar b u s$ tetraquark with quantum numbers $J^P = 1^+$. These values are reasonably consistent with results of other recent lattice-QCD studies of these systems \cite{Junnarkar:2018twb,Leskovec:2019ioa,Meinel:2022lzo,Hudspith:2023loy,Aoki:2023nzp}, but are in slight tension with earlier works \cite{Francis:2016hui,Mohanta:2020eed}. We have updated the summary plots for both tetraquark systems from Refs.~\cite{Leskovec:2019ioa,Meinel:2022lzo} and show them in Fig.~\ref{FIG:compare}. For the $\bar b \bar b u d$ system, also calculations \cite{Bicudo:2012qt,Brown:2012tm,Bicudo:2016ooe} based on static potentials from lattice QCD and the Born-Oppenheimer approximation are shown; these gave a somewhat smaller binding energy. The predictions from other approaches such as potential models, effective field theories, and QCD sum rules were taken from Refs.~\cite{Carlson:1987hh,SilvestreBrac:1993zem,Brink:1998as,Vijande:2003ki,Janc:2004qn,Vijande:2006jf,Navarra:2007yw,Ebert:2007rn,Zhang:2007mu,Lee:2009rt,Karliner:2017qjm,Eichten:2017ffp,Wang:2017uld,Park:2018wjk,Deng:2018kly,Wang:2018atz,Liu:2019stu,Tan:2020ldi,Lu:2020rog,Braaten:2020nwp,Faustov:2021hjs,Guo:2021yws,Dai:2022ulk,Kim:2022mpa,Chen:2022ros,Praszalowicz:2022sqx,Richard:2022fdc,Wu:2022gie,Maiani:2022qze,Song:2023izj,Liu:2023vrk}.

The main systematic uncertainties in our present results are associated with the lattice NRQCD action \cite{HPQCD:2011qwj}, in particular with the use of the tree-level value of the coefficient $c_4$ and the missing four-quark operators. The radiative corrections to $c_2$ and $c_4$, and the coefficients of the four-quark operators with four heavy quarks have in fact been computed to one loop in lattice perturbation theory in Refs.~\cite{Hammant:2013sca,Dowdall:2013jqa}. The systematic uncertainties could have been reduced by including these corrections, although four-quark operators with both heavy and light quarks would still be missing. A computation of the $\bar b \bar b u d$ and $\bar b \bar b u s$ tetraquark energies using a relativistic action for the $\bar b$ quarks is also a worthwhile future direction to complement and cross-check the results from this work and from Refs.\ \cite{Francis:2016hui,Junnarkar:2018twb,Leskovec:2019ioa,Mohanta:2020eed,Meinel:2022lzo,Hudspith:2023loy,Aoki:2023nzp}, which all used NRQCD.

Finally, we note that the calculations presented here only used scattering interpolating operators with zero relative momenta, leading to missing states in the energy range considered, especially for the larger volumes. A natural next step is to include also scattering operators with non-vanishing relative momenta. Their implementation is straightforward within our computational setup, and we have already included such operators in our recent investigation of the $\bar b \bar c u d$ systems \cite{Alexandrou:2023cqg}.

% ********************
% ********************
% ********************
% ********************
% ********************

\FloatBarrier

\section*{Acknowledgements}

We thank the MILC collaboration, and in particular Doug Toussaint, for providing the gauge-link ensembles.
We thank the developers of the QUDA \cite{Clark:2009wm,Babich:2011np,Clark:2016rdz} library, which was essential for the computations carried out in this project. We thank Luka Leskovec for collaboration on earlier related work. We thank Simone Bacchio for helpful discussions on the software used for the computations. We also acknowledge useful discussions with Ahmed Ali, Jozef Dudek, and Steve Sharpe.

C.A.\ acknowledges partial support by the project 3D-nucleon, ID number EXCELLENCE/0421/0043, co-financed by the European Regional Development Fund and the Republic of Cyprus through the Research and Innovation Foundation. J.F.\ received financial support by the German Research Foundation (DFG) research unit FOR5269 ``Future methods for studying confined gluons in QCD,'' by the PRACE Sixth Implementation Phase (PRACE-6IP) program (grant agreement No.~823767) and by the EuroHPC-JU project EuroCC (grant agreement No.~951740) of the European Commission. S.M.\ is supported by the U.S.\ Department of Energy, Office of Science, Office of High Energy Physics under Award Number D{E-S}{C0}009913.
M.P.\ and M.W.\ acknowledge support by the Deutsche Forschungsgemeinschaft (DFG, German Research Foundation) -- project number 457742095.
M.W.\ acknowledges support by the Heisenberg Programme of the Deutsche Forschungsgemeinschaft (DFG, German Research Foundation) -- project number 399217702.

We thank the Cyprus Institute for providing computational resources on Cyclone under the project IDs p054 and p147. This research also used resources of the National Energy Research Scientific Computing Center (NERSC), a U.S.\ Department of Energy Office of Science User Facility operated under Contract No.\ DE-AC02-05CH11231, and resources at the Texas Advanced Computing Center that are part of the Extreme Science and Engineering Discovery Environment (XSEDE), which is supported by National Science Foundation grant number ACI-1548562.
Calculations on the GOETHE-NHR and on the FUCHS-CSC  high-performance computers of the Frankfurt University were conducted for this research. We would like to thank HPC-Hessen, funded by the State Ministry of Higher Education, Research and the Arts, for programming advice.

% ********************
% ********************
% ********************
% ********************
% ********************

\appendix

\FloatBarrier

% ********************
% ********************
% ********************
% ********************
% ********************

\section{Additional plots for the other ensembles}

% ********************
% ********************
% ********************

\begin{figure}[htb]
	\centering
	\begin{minipage}{0.48\textwidth} $\quad \quad$ ensemble a15m310 \end{minipage} \hfill
	\begin{minipage}{0.48\textwidth} $\quad \quad$ ensemble a12m310 \end{minipage} \\
	\vspace{0.1cm}
	\includegraphics[width=0.48\textwidth]{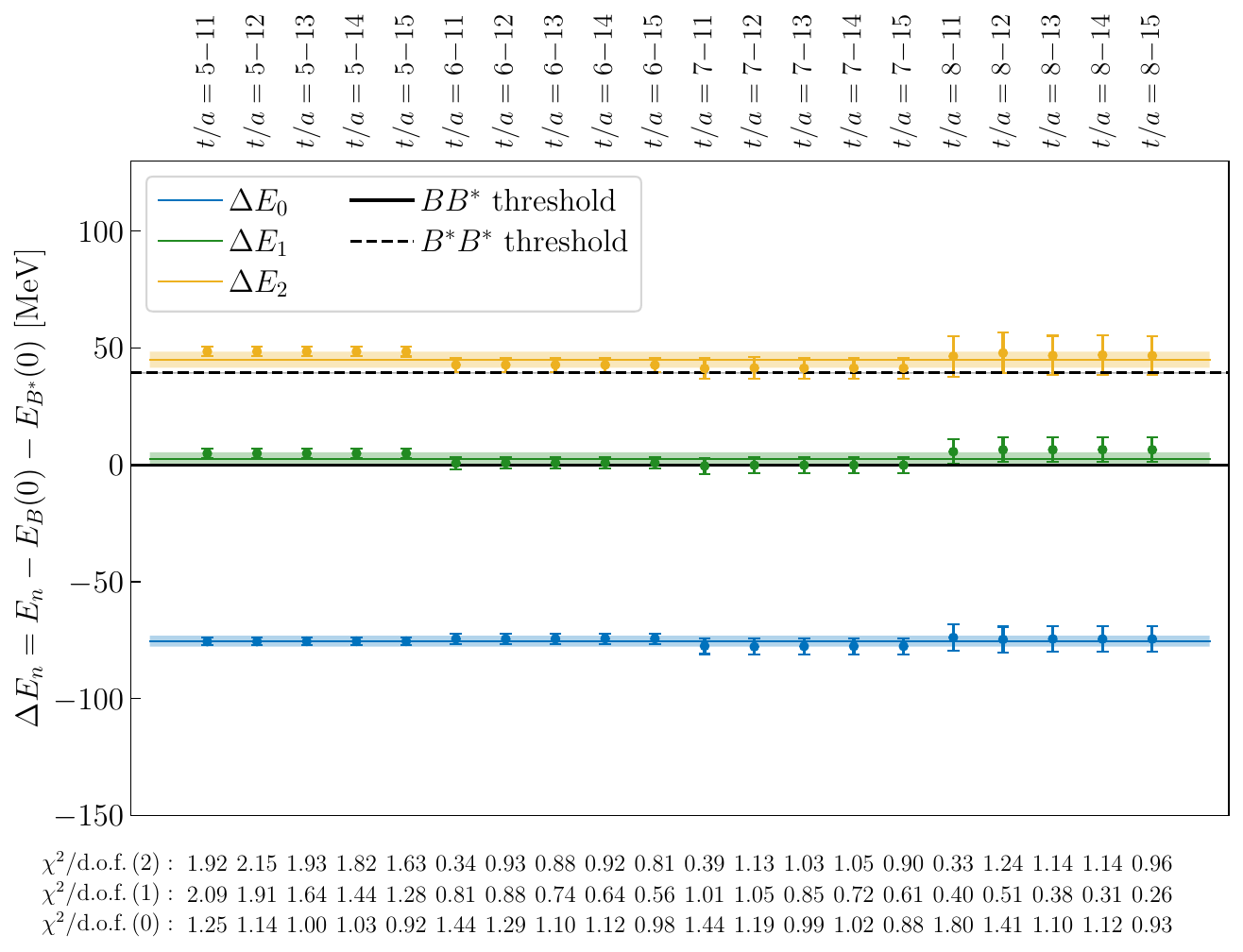} \hfill
	\includegraphics[width=0.48\textwidth]{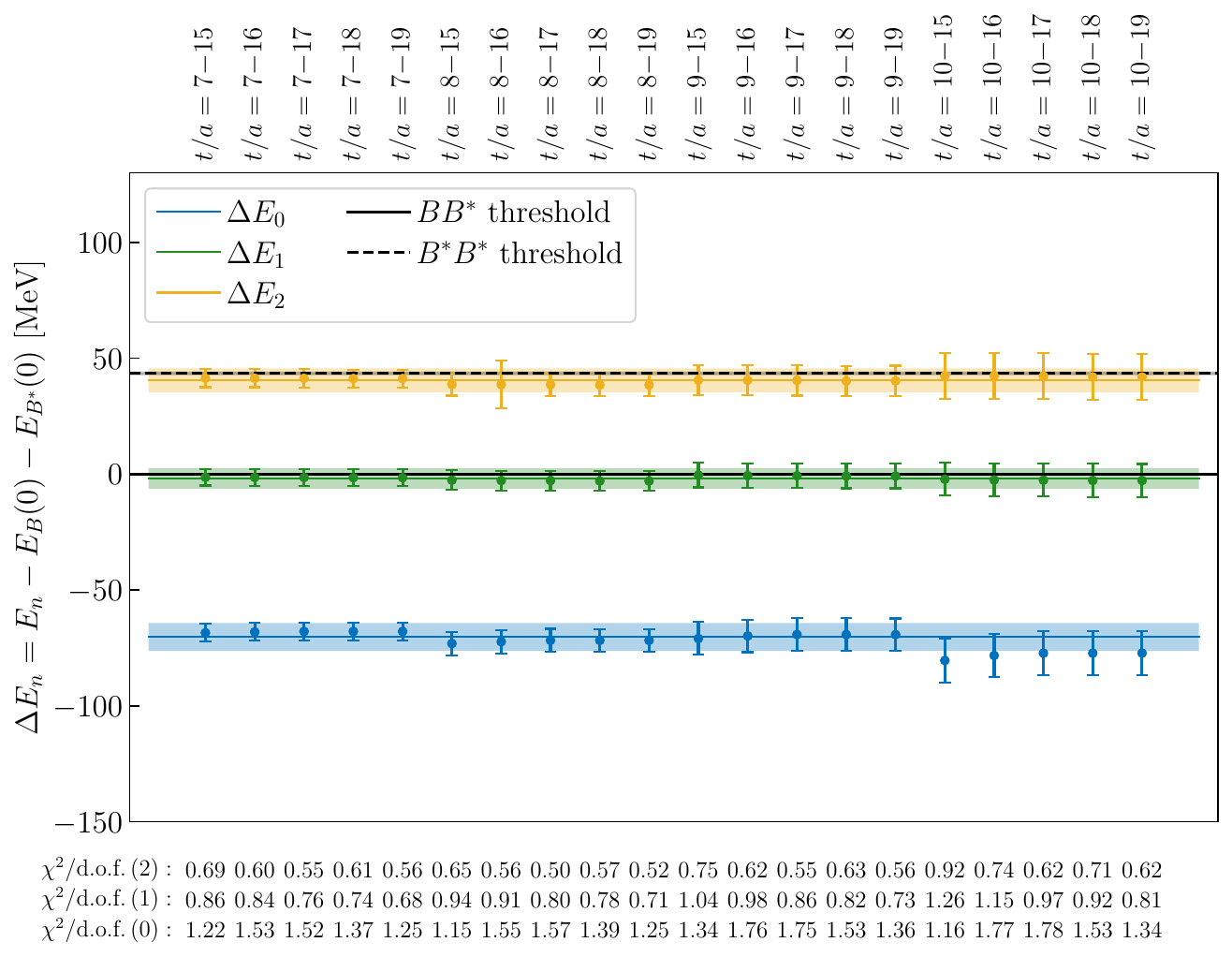} \\
	\vspace{0.2cm} \begin{minipage}{0.48\textwidth} $\quad \quad$ ensemble a12m220S \end{minipage} \hfill
	\begin{minipage}{0.48\textwidth} $\quad \quad$ ensemble a12m220 \end{minipage} \\
	\vspace{0.1cm}
	\includegraphics[width=0.48\textwidth]{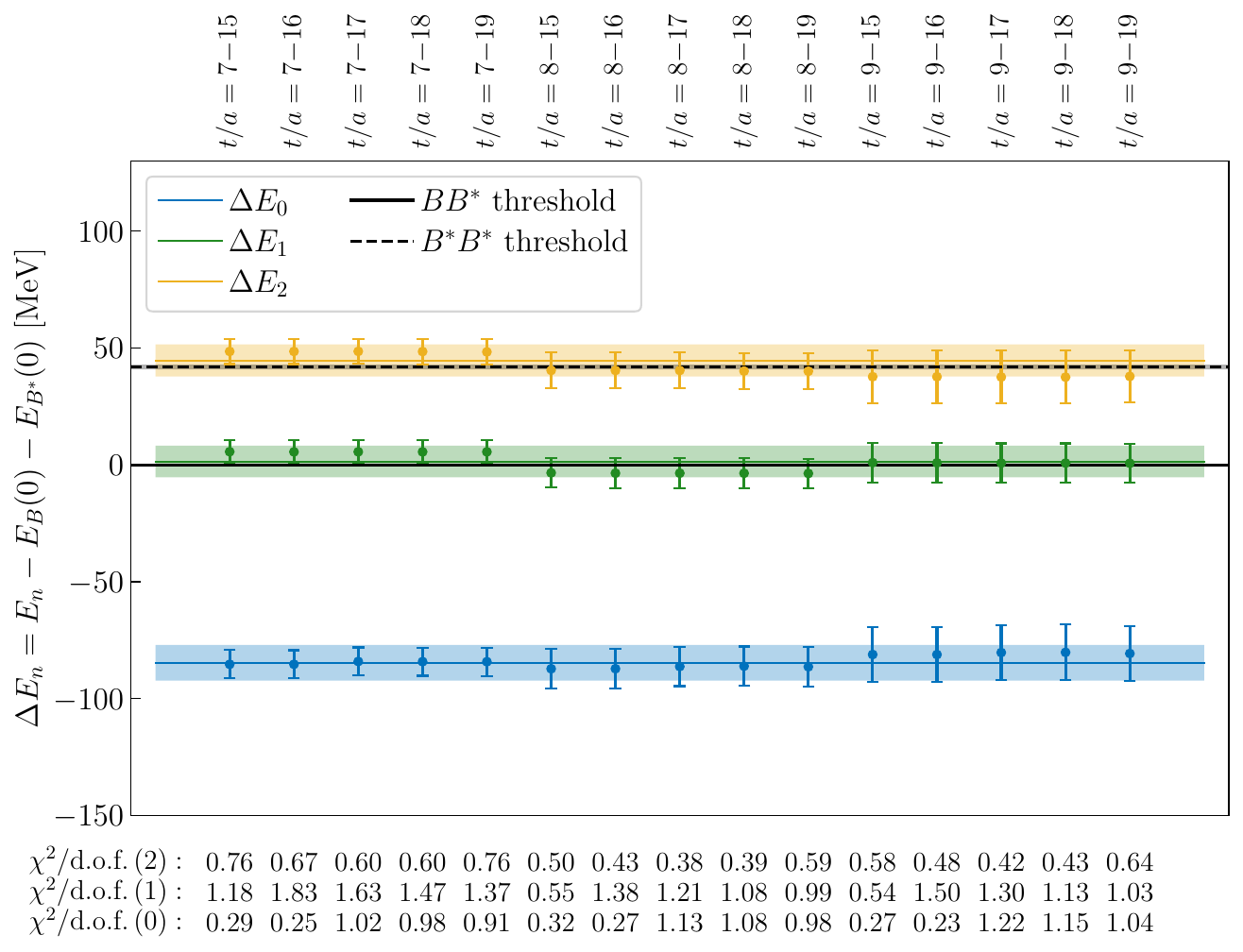} \hfill
	\includegraphics[width=0.48\textwidth]{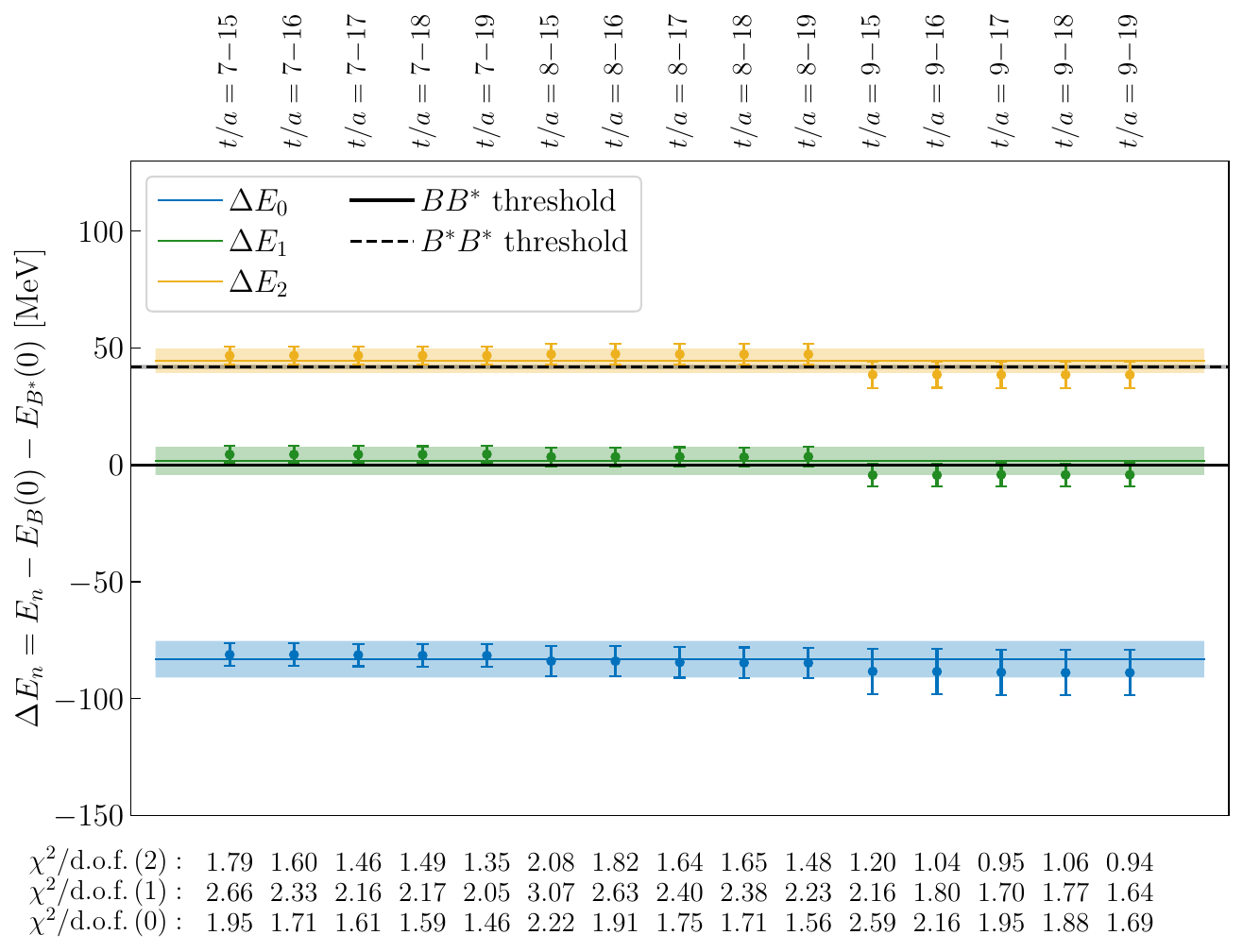} \\
	\vspace{0.2cm} \begin{minipage}{0.48\textwidth} $\quad \quad$ ensemble a09m310 \end{minipage} \hfill
	\begin{minipage}{0.48\textwidth} $\quad \quad$ ensemble a09m220 \end{minipage} \\
	\vspace{0.1cm}
	\includegraphics[width=0.48\textwidth]{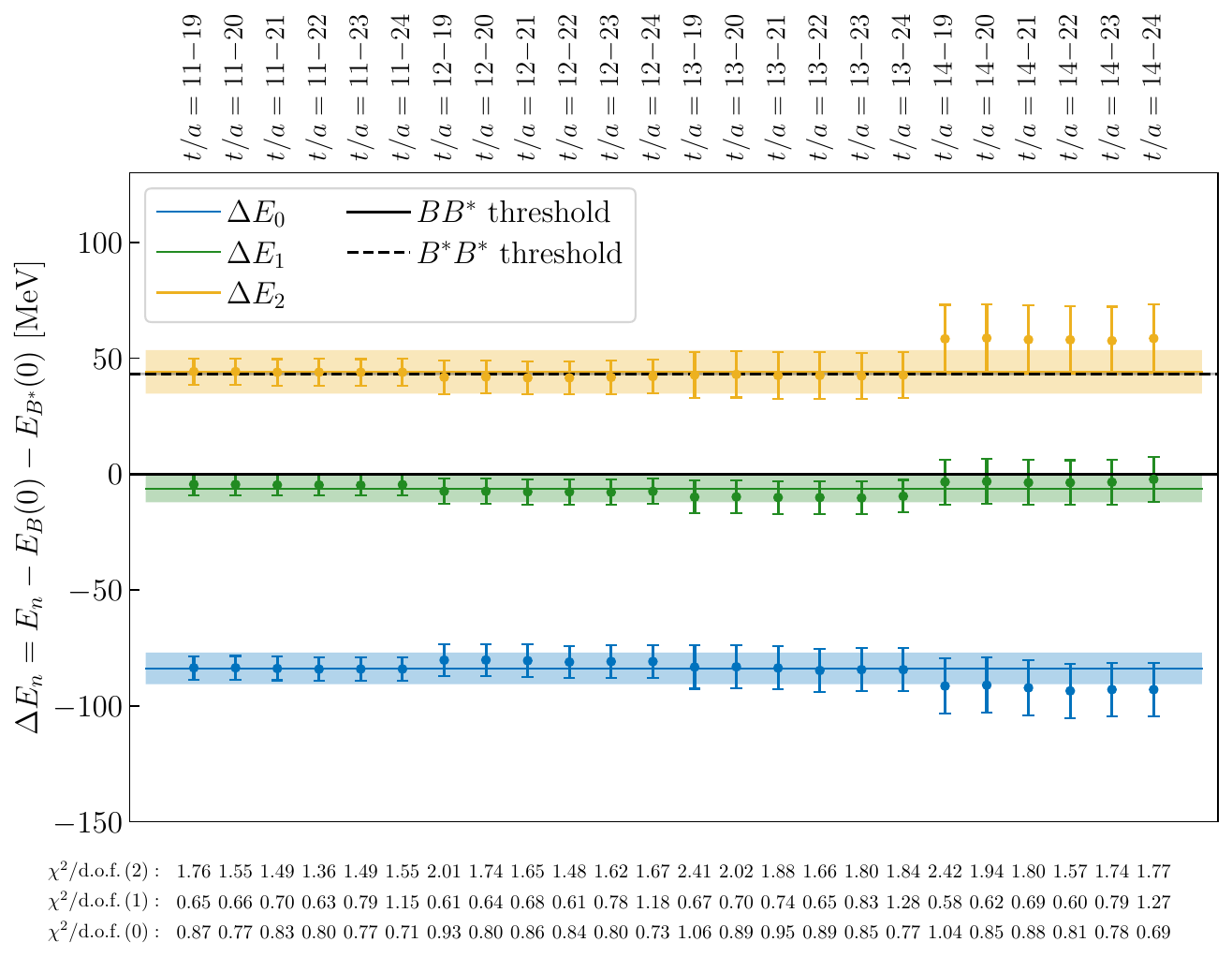} \hfill
	\includegraphics[width=0.48\textwidth]{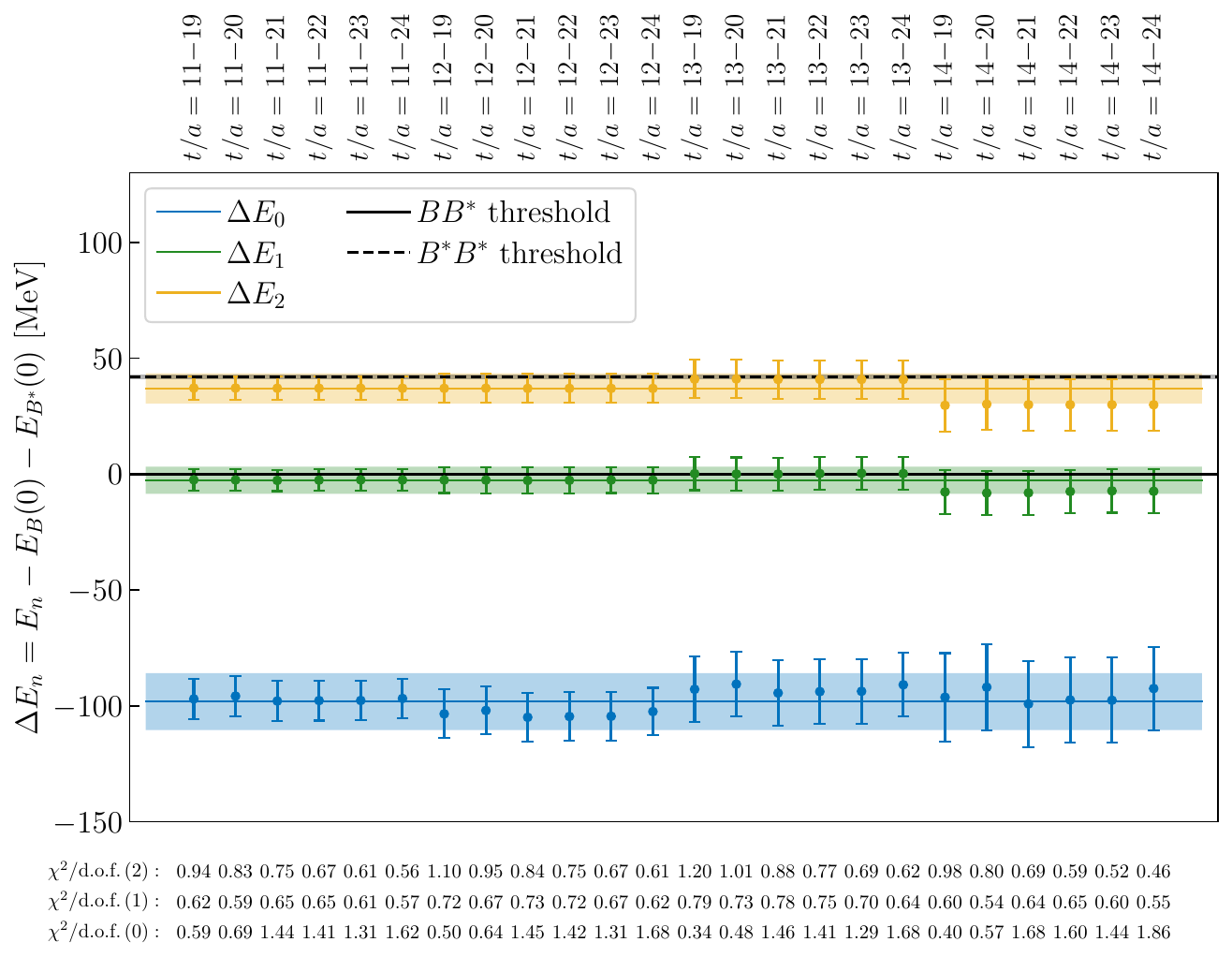}
	\caption{\label{fig:fitResults_bbud_appendix}Fit ranges and fit results for the energy levels for the $ \bbud $ system for the other ensembles.}
\end{figure}

% ********************
% ********************
% ********************

\begin{figure}[htb]
	\centering
	\begin{minipage}{0.48\textwidth} $\quad \quad$ ensemble a15m310 \end{minipage} \hfill
	\begin{minipage}{0.48\textwidth} $\quad \quad$ ensemble a12m310 \end{minipage} \\
	\vspace{0.1cm}
	\includegraphics[width=0.48\textwidth]{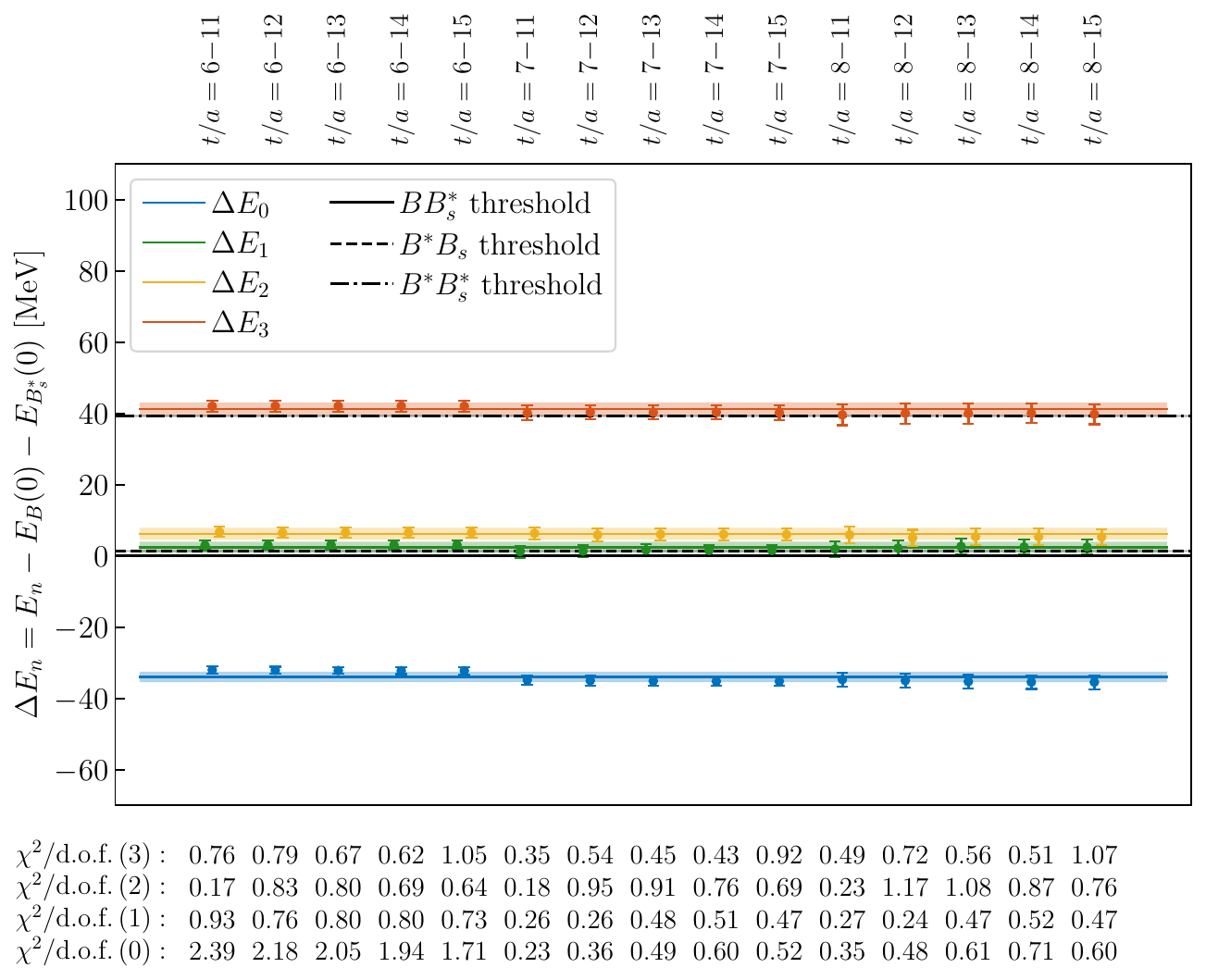} \hfill
	\includegraphics[width=0.48\textwidth]{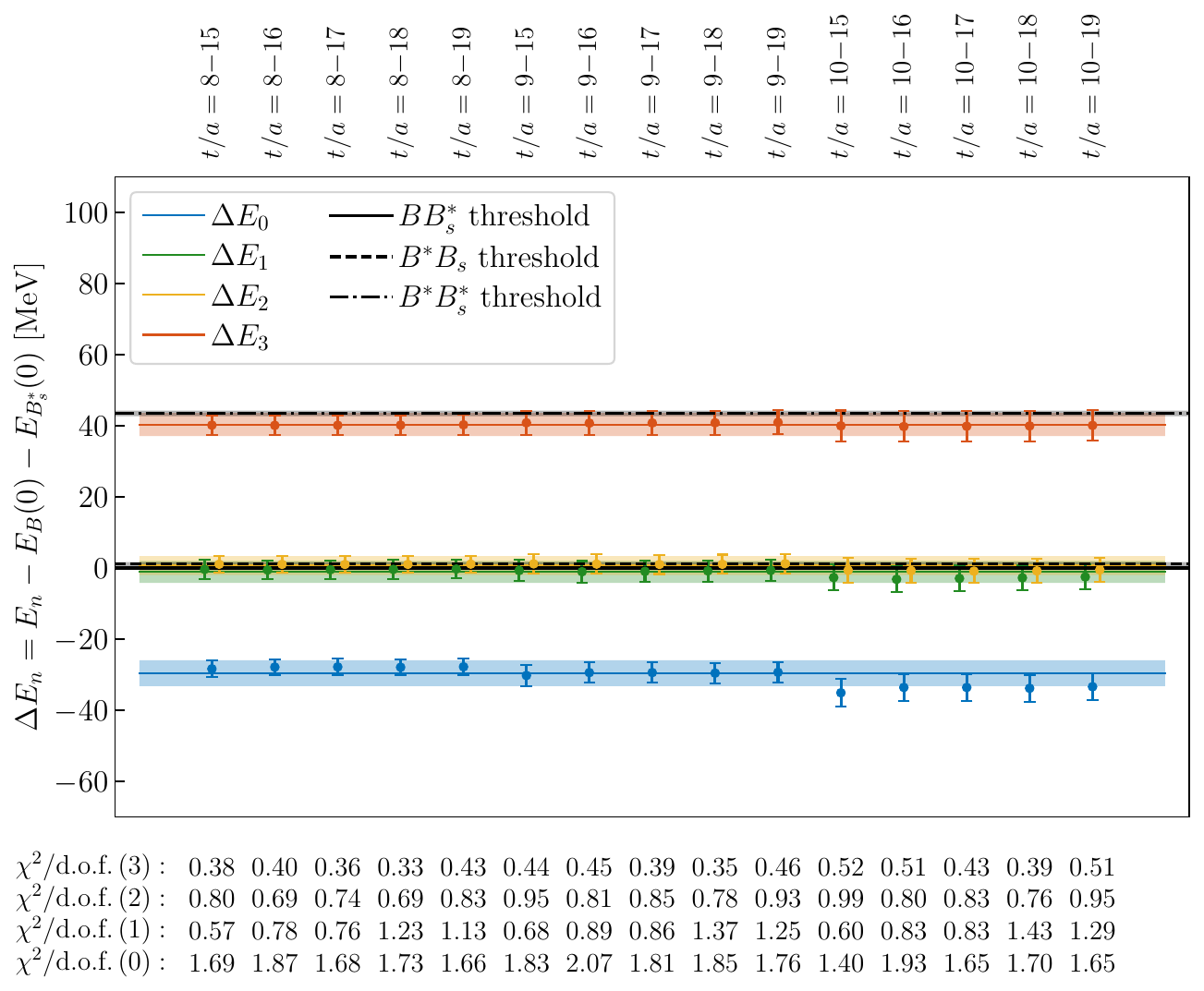} \\
	\vspace{0.2cm} \begin{minipage}{0.48\textwidth} $\quad \quad$ ensemble a12m220S \end{minipage} \hfill
	\begin{minipage}{0.48\textwidth} $\quad \quad$ ensemble a12m220 \end{minipage} \\
	\vspace{0.1cm}
	\includegraphics[width=0.48\textwidth]{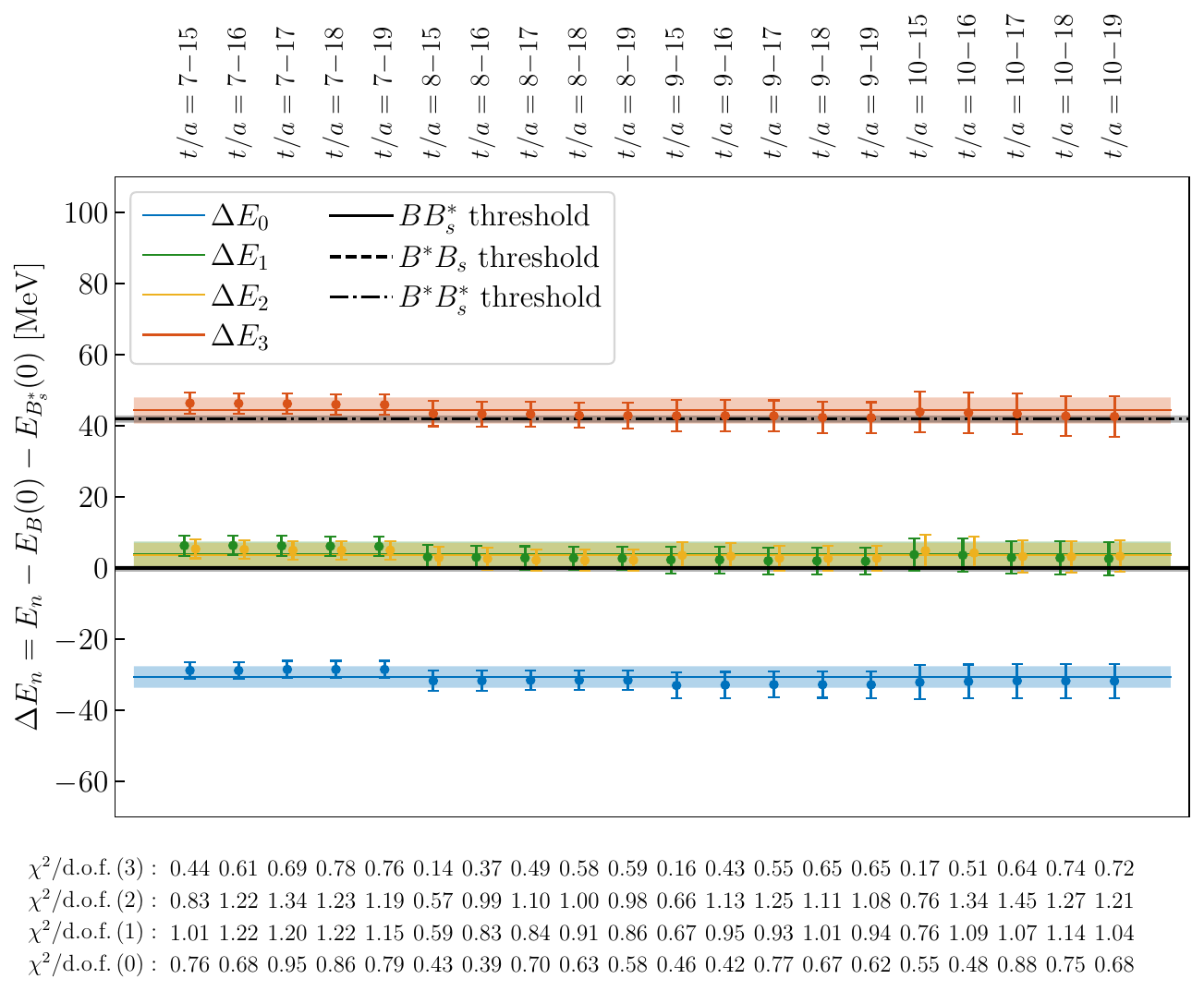} \hfill
	\includegraphics[width=0.48\textwidth]{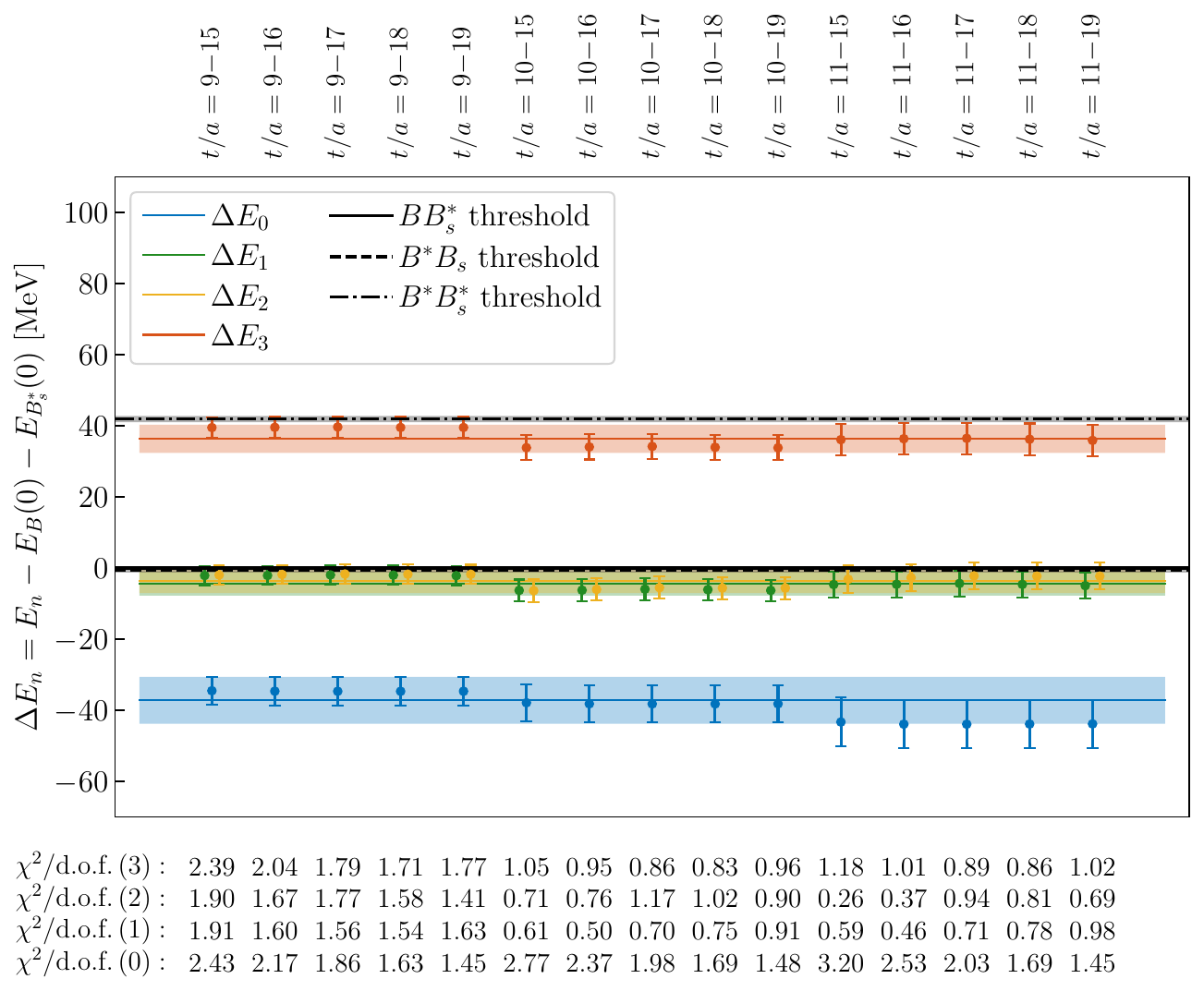} \\
	\vspace{0.2cm} \begin{minipage}{0.48\textwidth} $\quad \quad$ ensemble a09m310 \end{minipage} \hfill
	\begin{minipage}{0.48\textwidth} $\quad \quad$ ensemble a09m220 \end{minipage} \\
	\vspace{0.1cm}
	\includegraphics[width=0.48\textwidth]{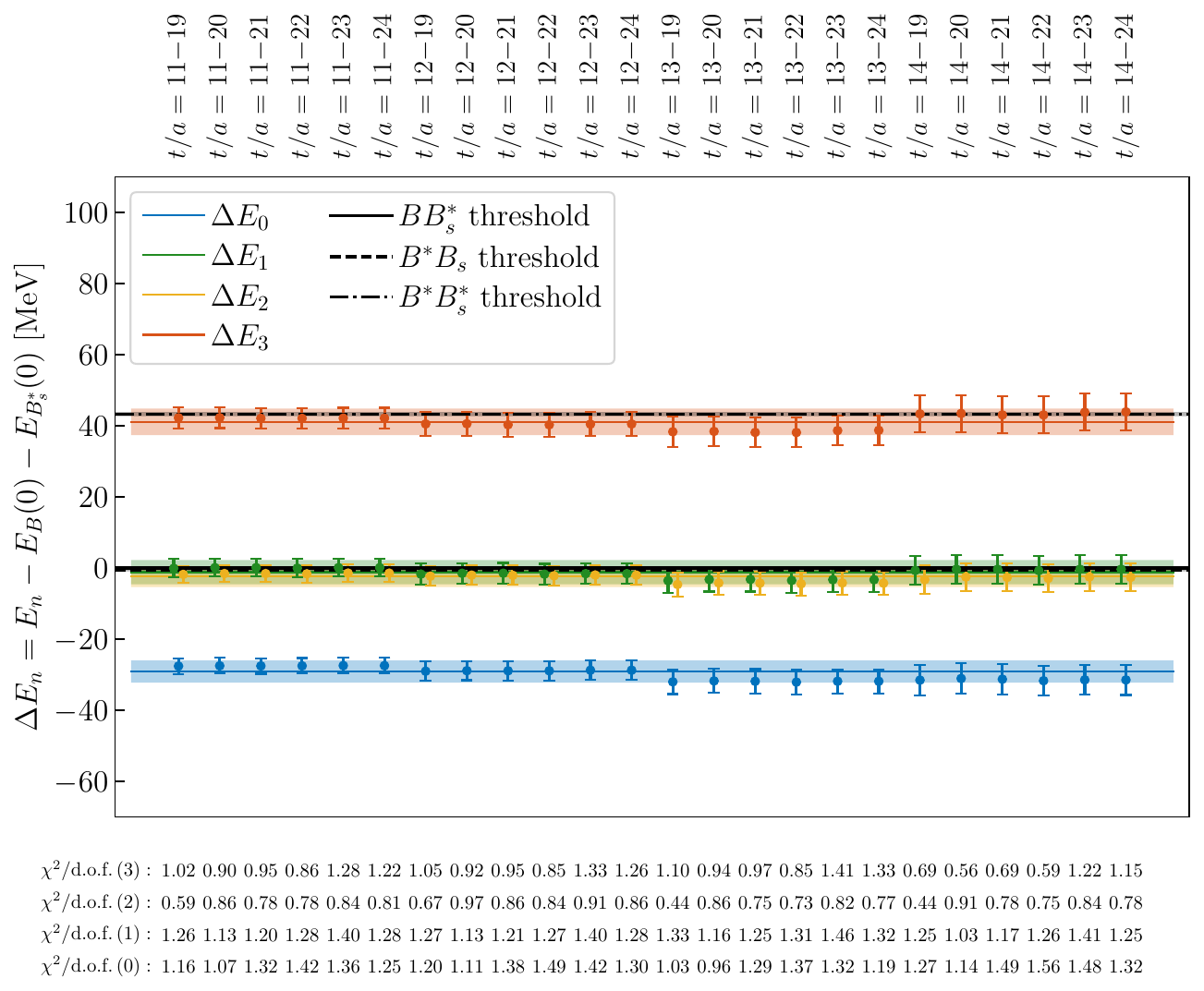} \hfill
	\includegraphics[width=0.48\textwidth]{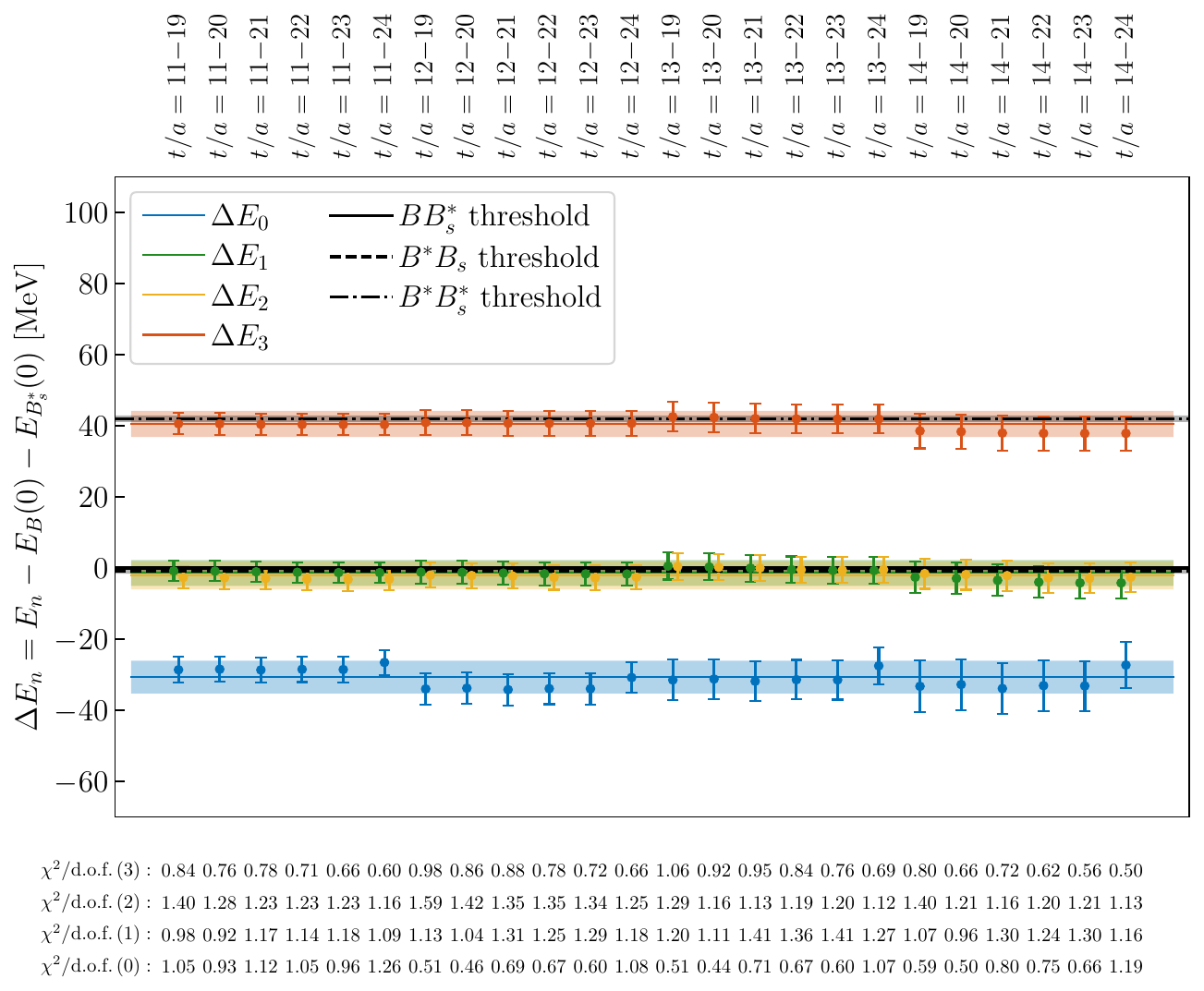}
	\caption{\label{fig:fitResults_bbus_appendix}Fit ranges and fit results for the energy levels for the $ \bbus $ system for the other ensembles.}
\end{figure}

\FloatBarrier

% ********************

\bibliographystyle{utphys-noitalics}
\bibliography{doubly-bottom-tetraquarks}

% ********************

\end{document}